\documentclass[prd,amsmath,aps,floats,amssymb, floatfix,superscriptaddress,nofootinbib,twocolumn,preprintnumbers]{revtex4-1}

\usepackage{lineno}

\usepackage{tabularx}
\usepackage{amsmath,amssymb,natbib,latexsym,times}
\usepackage[T1]{fontenc}
\usepackage{aecompl}
\usepackage{xspace}

\newcommand{\aap}{A\&A}
\newcommand{\apjs}{ApJS}
\newcommand{\aj}{A.J.}
\newcommand{\mnras}{MNRAS}

\newcommand{\jcap}{JCAP}
\newcommand{\apjl}{ApJ}

\usepackage{verbatim}
\usepackage[usenames,dvipsnames]{xcolor}
\usepackage{tabulary}
\usepackage{tabularx}
\usepackage{todonotes}
\usepackage{hyperref}
\usepackage{ulem}
\usepackage{bm}

\newcommand{\healpix}{\code{HEALPix}\xspace}

\newcommand{\photoz}{photo-$z$\xspace}

\newcommand{\der}{\mathrm{d}}

\newcommand\lcdm{$\Lambda$CDM\xspace}

\newcommand*\justify{%
  \fontdimen2\font=0.4em
  \fontdimen3\font=0.2em
  \fontdimen4\font=0.1em
  \fontdimen7\font=0.1em
  \hyphenchar\font=`\-
}

\newcommand\code[1]{\texttt{\small\justify #1}}

\newcommand{\DNF}{\textsc{DNF}\xspace}

\newcommand{\ZMEAN}{\code{Z\_MEAN}\xspace}
\newcommand{\ZMC}{\code{Z\_MC}\xspace}

\newcommand{\ngmix}{\textsc{ngmix}\xspace}

\newcommand\cosmolike{{\textsc{CosmoLike} }}
\newcommand{\namaster}{\textsc{NaMASTER}\xspace}

\newcommand{\gold}{\code{Y3\,GOLD}\xspace}
\newcommand{\sample}{{\rm {BAO\,sample}}\xspace}

\newcommand{\var}[1]{\ensuremath{\texttt{\MakeUppercase{#1}}}\xspace}

\newcommand\be{\begin{equation}}
\newcommand\ee{\end{equation}}
\def\bea{\begin{eqnarray}}
\def\eea{\end{eqnarray}}

\newcommand\mpc{\,{\it h}^{-1}\,{\rm Mpc}}
\newcommand\invmpc{\,{\it h}\,{\rm Mpc}^{-1}}

\newcommand\T{\rule{0pt}{2.6ex}}       
\newcommand\B{\rule[-1.2ex]{0pt}{0pt}} 

\newcommand\red[1]{\textcolor{Black}{#1}}

\allowdisplaybreaks

\begin{document}
\title{Dark Energy Survey Year 3 Results: A 2.7$\%$ measurement of Baryon Acoustic Oscillation distance scale at redshift 0.835}

\author{T.~M.~C.~Abbott}
\affiliation{Cerro Tololo Inter-American Observatory, NSF's National Optical-Infrared Astronomy Research Laboratory, Casilla 603, La Serena, Chile}
\author{M.~Aguena}
\affiliation{Laborat\'orio Interinstitucional de e-Astronomia - LIneA, Rua Gal. Jos\'e Cristino 77, Rio de Janeiro, RJ - 20921-400, Brazil}
\author{S.~Allam}
\affiliation{Fermi National Accelerator Laboratory, P. O. Box 500, Batavia, IL 60510, USA}
\author{A. Amon}
\affiliation{Kavli Institute for Particle Astrophysics \& Cosmology, P. O. Box 2450, Stanford University, Stanford, CA 94305, USA}
\author{F.~Andrade-Oliveira}
\affiliation{Instituto de F\'{i}sica Te\'orica, Universidade Estadual Paulista, S\~ao Paulo, Brazil}
\affiliation{Laborat\'orio Interinstitucional de e-Astronomia - LIneA, Rua Gal. Jos\'e Cristino 77, Rio de Janeiro, RJ - 20921-400, Brazil}
\author{J.~Asorey}
\affiliation{Centro de Investigaciones Energ\'eticas, Medioambientales y Tecnol\'ogicas (CIEMAT), Madrid, Spain}
\author{S.~Avila}
\affiliation{Instituto de Fisica Teorica UAM/CSIC, Universidad Autonoma de Madrid, 28049 Madrid, Spain}
\author{G.~M.~Bernstein}
\affiliation{Department of Physics and Astronomy, University of Pennsylvania, Philadelphia, PA 19104, USA}
\author{E.~Bertin}
\affiliation{CNRS, UMR 7095, Institut d'Astrophysique de Paris, F-75014, Paris, France}
\affiliation{Sorbonne Universit\'es, UPMC Univ Paris 06, UMR 7095, Institut d'Astrophysique de Paris, F-75014, Paris, France}
\author{A.~Brandao-Souza}
\affiliation{Instituto de F\'isica Gleb Wataghin, Universidade Estadual de Campinas, 13083-859, Campinas, SP, Brazil}
\affiliation{Laborat\'orio Interinstitucional de e-Astronomia - LIneA, Rua Gal. Jos\'e Cristino 77, Rio de Janeiro, RJ - 20921-400, Brazil}
% ...

\collaboration{DES Collaboration}

\noaffiliation
\date{\today}
\label{firstpage}
\begin{abstract}
We present angular diameter  measurements obtained by measuring the position of Baryon Acoustic Oscillations (BAO) in an optimised sample of galaxies from \red{the first three years} of Dark Energy Survey data (DES Y3). The sample consists of  {7 million} galaxies distributed over a footprint of 4100 deg$^2$ with $0.6 < z_{\rm photo} < 1.1$ and a typical redshift uncertainty of $0.03(1+z)$. The sample selection is the same as in the BAO measurement with the first year of DES data, but the analysis presented here uses three times the area, extends to higher redshift and makes a number of improvements, \red{including a fully analytical BAO template, the use of covariances from both theory and simulations, and an extensive pre-unblinding protocol}. We used two different statistics: angular correlation function and power spectrum, and validate our pipeline with an ensemble of over 1500 realistic simulations. Both statistics yield compatible results.
We combine the likelihoods derived \red{from angular correlations and spherical harmonics} to constrain the ratio of comoving angular diameter distance $D_M$ at the effective redshift of our sample to the sound horizon scale at the drag epoch. We obtain $D_M(z_{\rm eff}=0.835)/r_{\rm d} = 18.92 \pm 0.51$, which is \red{consistent with, but smaller than,} the Planck prediction assuming flat \lcdm, at the level of $2.3 \sigma$. The analysis was performed blind and is robust to changes in a number of analysis choices. It represents the most precise BAO distance measurement from imaging data to date, and is competitive with the latest transverse ones from spectroscopic samples at $z>0.75$. When combined with DES 3x2pt + SNIa, they lead to improvements in $H_0$ and $\Omega_m$ constraints by $\sim 20\%$.
\end{abstract}


\preprint{DES-2021-0651}
\preprint{FERMILAB-PUB-15-849-PPD}

\maketitle

\section{Introduction}

The scientific knowledge of the Universe changed in the 1990s \red{with the discovery
of its accelerated expansion} \citep{1998AJ....116.1009R,1999ApJ...517..565P}. This 
remarkable discovery opened the door to the current cosmological standard 
model, \lcdm. The model is very well established, since it is based on a very large set of 
independent observations that it can explain with high accuracy. However, it has 
shocking consequences. Only a small fraction (around 5\%) of the content of our 
Universe is ordinary matter. The other 95\% is composed of exotic entities called 
dark matter and dark energy,  that have not been detected yet in laboratories. \lcdm
describes the dominant component of the matter-energy content of the Universe, dark 
energy, as a cosmological constant $\Lambda$, a description that is in agreement with all the 
current cosmological observations~\citep{2020PTEP.2020h3C01P}. However, the value of $\Lambda$ itself is 
not compatible with the Standard Model of particle physics \citep{RevModPhys.61.1, Padmanabhan:2002ji}. The current measurements of 
dark energy properties \red{are not yet precise enough to distinguish between possible}
explanations, like the incompleteness of General Relativity to describe gravity at cosmological scales or 
the existence of some mysterious fluid with negative pressure filling the whole 
Universe, are still possible.

Several observational probes are used to study the nature of dark energy. Among 
them, the measurement of the evolution with redshift of the angular diameter 
distance and the Hubble distance, using the scale of the Baryon Acoustic Oscillations (BAO) as a standard ruler, is 
one of the most robust, since it is insensitive to systematic uncertainties related to 
the astrophysical properties of the tracers (galaxies, quasars or the Lyman-$\alpha$ forest). In addition, the physics that causes BAO is 
well understood, which allows very precise measurements with the current cosmological
surveys. The BAO signal was first detected in 2005 by both 
the Sloan Digital Sky Survey (SDSS)  \citep{2005ApJ...633..560E} and the 2-degree Field Galaxy Redshift Survey (2dFGRS) \citep{2001MNRAS.327.1297P,2005MNRAS.362..505C}. Today there are many detections at different 
redshifts from SDSS I, II, III and IV \citep{2007MNRAS.381.1053P,2009MNRAS.399.1663G,2010MNRAS.401.2148P,2015MNRAS.449..835R,2017MNRAS.470.2617A,2018MNRAS.473.4773A,Ly-alpha_BOSS,Ly-alpha_BOSS2,Ly-alpha_cross, 2015A&A...574A..59D,2017A&A...603A..12B,2020ApJ...901..153D,2021PhRvD.103h3533A}, the
6-degree Field Galaxy Survey (6dFGS) \citep{2011MNRAS.416.3017B} and 
WiggleZ \citep{2011MNRAS.415.2892B,2011MNRAS.418.1707B,2014MNRAS.441.3524K,2017MNRAS.464.4807H}, mapping the redshift evolution of the angular diameter distance.

The Dark Energy Survey (DES, \cite{2005IJMPA..20.3121F,2016MNRAS.460.1270D}) is a photometric galaxy survey that probes the physical nature of dark energy by means of several independent and complementary probes.
One of these 
probes is the precise study of the spatial distribution of galaxies, and in particular, the 
BAO standard ruler. Since DES is a photometric survey, its precision in the measurement of 
redshifts is limited. However, a precise determination of the evolution of the angular distance 
with redshift is possible through the measurement of angular correlation functions or angular 
power spectra, as was already probed with the DES Y1 data \citep{2019MNRAS.483.4866A}. Other 
measurements of the BAO signal in various photometric data samples have been presented in Refs. 
\cite{2007MNRAS.378..852P,2009ApJ...692..265E,2010MNRAS.401.2477H, 2011MNRAS.411..277S,2011MNRAS.417.2577C,2012ApJ...761...13S,2012MNRAS.419.1689C} using a variety of methodologies.

In this paper, a new determination of the BAO scale is presented. The measurement uses the
imaging data from DES Y3, described in \cite{y3-gold}, to measure the angular 
diameter distance to red galaxies that have been specially selected \citep{y3-baosample} with 
photometric redshifts $0.6 < z_{\rm photo} < 1.1$. DES has imaged a 5,000 deg$^2$ footprint of 
the Southern hemisphere using five passbands, $grizY$, collecting the properties of over 
300 million galaxies. Here, we use 7 million galaxies over 4100 deg$^2$, that were color and magnitude 
selected to balance trade-offs in BAO measurements between the redshift precision and the 
number density. We use these data, supported by 1952 mock realizations of our sample, to 
measure the BAO scale at an effective redshift of $0.835$. 

The measurement we present is supported by a series of companion papers. In \cite{y3-baosample} 
we present all the details of the selection of the galaxy sample, that was optimised for 
$z > 0.6$ BAO measurements, the tests of its basic properties, the mitigation of observational 
systematic effects and the photometric redshift validations, \red{extending what was done 
in \cite{DESY1baosample}}.  In \cite{y3-baomocks} we describe the mock catalogues 
we have developed to test the measurement methods. These mocks match the spatial properties of
the data sample and the \photoz resolution.

The structure of this paper is as follows. Section~\ref{sec:data} summarizes the data we used, including all of its basic properties and the process we used to mitigate systematic effects. Section~\ref{sec:simulations} presents the different sets of simulations used to validate the analysis. Section~\ref{sec:analysis} describes the analysis methodology, including the techniques to measure the clustering of galaxies (both angular correlation function and angular power spectrum), the estimation of the covariance matrices and the production of the templates that have been used to fit the BAO scale. In addition, it also presents how the BAO scale is extracted from the data. Section~\ref{sec:mocktest} present the validation of the entire analysis using simulations. Section~\ref{sec:blindtest} describes the full set of pre-unblinding tests that we defined as a requirement to be passed before revealing the final results.  Section~\ref{sec:results} presents our results, starting with the clustering measurements and the resulting BAO scale followed by the distance measurement derived from this scale. Finally, Section~\ref{sec:cosmo} is devoted to the cosmological implications of this measurement, including a comparison to predictions of the flat \lcdm model and other BAO scale distance measurements. Our conclusions are presented in Section~\ref{sec:conclusions}.

The fiducial cosmology we use to quote our primary results is a flat \lcdm model with $\Omega_m = 0.31$, $h = 0.676$, $n_s=0.97$ and $\sigma_8=0.83$; consistent with \cite{2020A&A...641A...6P} (we denote this as {\tt Planck} hereafter). However our mocks, and therefore all the statistical and systematics tests to validate our methodology, were carried out using $\Omega_m = 0.25$ and $h = 0.7$ (denoted as {\tt MICE}  in what follows).  This cosmology matches the one of the MICE N-Body simulations \cite{MICE1,MICE2,MICE3}, which was used to calibrate the mock galaxy 
samples themselves. We demonstrate that our results are not sensitive to this choice.

It is important to remark that the whole analysis was performed blinded. The sample selection cuts and the estimation of photometric redshifts and redshift distributions, the  treatment of observational systematics, and the analysis choices were defined and completed a priori. In addition, a detailed set of tests to be passed before unblinding the results was put in place. Only when the analysis passed this predefined criteria, we unblinded the measurement of the BAO scale.

\section{Dark Energy Survey Data}
\label{sec:data}

\subsection{DES Y3-GOLD catalogue}

The Dark Energy Survey (DES) observed for six years using the Dark Energy Camera (DECam \cite{2015AJ....150..150F}) at the Blanco 4m telescope at the Cerro Tololo Inter-American Observatory (CTIO) in Chile. 
The survey covered 5000 sq. deg. in $grizY$ bandpasses to approximately 10 overlapping dithered exposures in each filter.
We utilize data taken during the first three years of DES operations (DES Y3), which made up DES Data Release 1 (DR1 \cite{2018ApJS..239...18A}). 
This analysis covers the full 5000 sq. deg. survey footprint for the first time, but at approximately half the full-survey depth. 
The data is processed, calibrated, and coadded to produce a photometric data set of 390 million objects that is further refined to a `Gold' sample for cosmological use \cite{y3-gold}. 
The \gold sample includes cuts on minimal image depth and quality, additional calibration and deblending, and quality flags to identify problematic photometry and regions of the sky with substantial photometric degradation (e.g., around bright stars). The \gold sample extends to a 10$\sigma$ limiting magnitude of 23 in $i$-band and it is the basis for the definition of the BAO sample.

\subsection{BAO Sample}
\label{sec:baosample}

We select \red{a subsample of} red galaxies from the \gold sample \cite{y3-gold} following the same color selection as in Y1 \citep{DESY1baosample}, designed to balance the sample density with the \photoz precision above redshifts greater than $0.5$. The sample covers 4108.47 ${\rm deg}^{2}$, almost three times larger than the Y1 \sample, and comprises $7,031,993$ galaxies in the redshift range $0.6< z <1.1$, up to a magnitude limit of $i<22.3 \ (AB, 10\sigma$). Full details about the sample selection and characterization is found in \cite{y3-baosample}, here we summarize the main properties.

Improvements in data reduction and processing between Y1 and Y3 increased the DES detection efficiency, which translated into a higher number density of sources, even though the number of exposures per pointing in the footprint is in average the same. We take full advantage of this optimization for the \sample and extend the analysis from $z_{\rm photo} = 1$ to $z_{\rm photo} = 1.1$ and to a fainter magnitude limit with respect to Y1 \red{(22.3 instead of 22, in the $i$-band)}. Our forecasts showed that this extension of the sample meant a $10\%$ gain in the precision of the combined BAO distance measurement.

During the selection process we used \ngmix\footnote{\href{https://github.com/esheldon/ngmix}{https://github.com/esheldon/ngmix}} \var{SOF} magnitudes \red{(hereafter referred to as \var{SOF})} with chromatic corrections and dereddened using SED-dependent extinction corrections. These magnitudes are defined in \cite{y3-gold}. We also use the morphological classification based on \var{SOF} photometry to select secure galaxies. These choices are common to all DES Y3 cosmological results.

We start by applying the same color cut as in Y1 to the Gold sample to select red galaxies. Despite the changes in photometry from Y1 and Y3, the color selection still isolates galaxies at $z > 0.5$, as attested in \citep{y3-baosample}. The primary selection includes the aforementioned color cut, a magnitude cut as a function of photometric redshift $z_{\rm photo}$ (to remove fainter galaxies at lower redshifts) and the redshift range. The cuts applied are:
\begin{eqnarray}
&&(i_\var{SOF}-z_\var{SOF}) + 2.0 (r_\var{SOF}-i_\var{SOF}) > 1.7,  \label{eq:colorcut} \\ 
 &&i_\var{SOF} < 19 + 3.0 \, z_{\rm photo},  \label{eq:magcut} \\
 &&0.6<z_{\rm photo}<1.1, \label{eq:zrange} 
\end{eqnarray} 
where $z_{\rm photo}$ is the photometric redshift estimate, which we describe in detail in Sec.~\ref{sec:photo-z}. 
In addition we impose a bright magnitude cut $17.5 < i_\var{SOF}$ to remove bright contaminant objects such as binary stars.
Stellar contamination is mitigated with the galaxy and star classifier \var{EXTENDED\_CLASS\_MASH\_SOF} from the \gold catalogue. 
Likewise, we remove sources that are flagged as suspicious or with corrupted photometry (see \cite{y3-baosample}).

A summary of the \sample properties is given in Table \ref{tab:sample}, and an in-depth discussion of the selection process and flags can be found in \cite{y3-baosample}.

\begin{table}
\centering
\caption{Main properties of the \sample as a function of tomographic bin: mean redshift, number of galaxies, mean photo-$z$ accuracy (in units of $(1+z)$) and $68\%$ width of the true redshift distribution estimates, see Sec.~\ref{sec:photo-z}. The latter are found by stacking spectroscopic redshifts from a matched sample with the VIPERS dataset. The sample covers 4108.47 deg$^{2}$. Full details can be found  in \cite{y3-baosample}.} 
\begin{tabular}{lccccc} \\
\hline
\hline
	$z_{\rm photo}$ & $\bar{z}$ & $N_{\rm gal}$ & $\sigma_{68}$ & $W_{68}$ \\
\hline
	$0.6 < z < 0.7$ \ \ \ & 0.65  \ \ \ & 1478178   \ \ \ & 0.021  \ \ \  & 0.045 \\
	$0.7 < z < 0.8$ \ \ \  & 0.74  \ \ \ & 1632805   \ \ \ & 0.025  \ \ \ & 0.052 \\
	$0.8 < z < 0.9$ \ \ \  & 0.84  \ \ \ & 1727646   \ \ \ & 0.029  \ \ \ & 0.063\\
	$0.9 < z < 1.0$ \ \ \  & 0.93  \ \ \ & 1315604   \ \ \ & 0.030  \ \ \ & 0.063 \\
	$1.0 < z < 1.1$ \ \ \  & 1.02  \ \ \ & 877760    \ \ \ & 0.040  \ \ \ & 0.081 \\
\hline
\hline
\label{tab:sample}
\end{tabular}
\end{table}

\subsection{Photometric Redshifts}
\label{sec:photo-z}

We measure the BAO scale in tomographic redshift bins of width $\Delta z=0.1$ from $0.6$ to $1.1$. In order to assign galaxies to each redshift bin we use the \photoz estimate given by the \red{Directional Neighborhood Fitting (\DNF)} algorithm \cite{DNF}, which was trained using \var{SOF} $griz$ fluxes on a large training sample. In this work, this reference dataset includes $\sim 2.2 \times 10^5$ spectra matched to DES objects from 24 different spectroscopic catalogues, in particular SDSS DR14 \cite{sdssdr14} and the OzDES program \cite{ozdes:2017}. \DNF performs a nearest-neighbors fit to the hyper-plane in color and magnitude space to this training set, and predicts the best \photoz estimate (called \ZMEAN in the DES catalogues), as well as the redshift of the closest neighbor (\ZMC) and the full PDF distribution. We use \ZMEAN to assign galaxies to each tomographic bin. 

In order to estimate the true redshift distribution $n(z)$ in each tomographic bin we construct a matched sample \cite{y3-baosample}, within one arcsec apertures, with the second public data release (PDR2) \cite{vipers_pdr2} from ``VIMOS Public Extragalactic Redshift Survey'' (VIPERS) \cite{vipers}. VIPERS was designed to be a complete galaxy sample up to $i<22.5$ for redshifts above $0.5$ and covers $16.32$ deg$^2$ in the DES footprint, containing 74591 \sample galaxies in the matched catalogue. Each of these galaxies is weighted to account for target selection, colour selection and spectroscopic efficiency. We stack the spectroscopic redshifts $z_{\rm VIPERS}$ from the matched catalogue to estimate our five $n(z)$'s\footnote{For consistency, we had removed VIPERS from the DNF training sample.}. In \cite{y3-baosample} we also use the stacking of the \DNF \ZMC or \DNF PDF's to validate the redshift distributions of the \sample. 

Table \ref{tab:sample} contains estimates of the photometric redshift accuracy per galaxy, $\sigma_{68}$, defined as the half width of the interval containing the median $68\%$ of values in the distribution of $(\ZMEAN-z_{\rm VIPERS})/(1+z_{\rm VIPERS})$. It also shows an estimate of the width of the individual $n(z)$'s, $W_{68}$, similarly defined as the $68\%$ confidence region of the stacking of $z_{\rm VIPERS}$. It also contains the mean of each $n(z)$, $\bar{z}$, that matches well the geometric mean of the corresponding bin edges, except at the last bin where the distribution is a bit skewed towards low redshifts.

\subsection{Angular Mask}
\label{sec:mask}

\begin{figure}
\includegraphics[width=84mm]{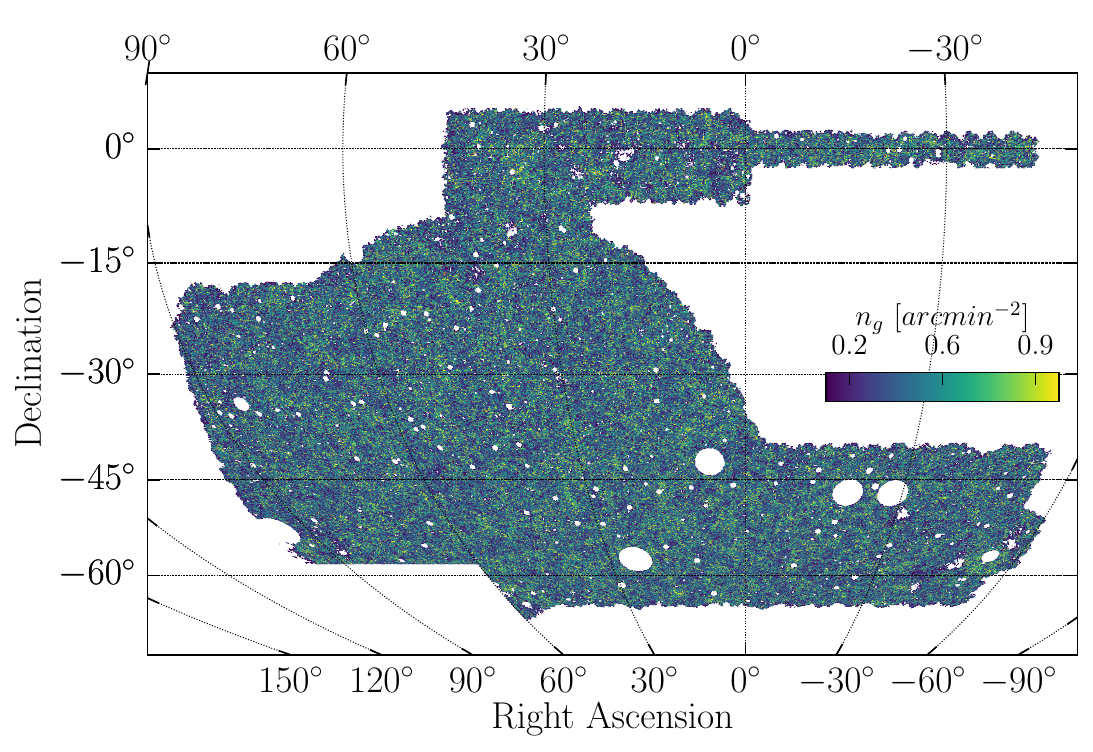}  
\caption{\sample projected density field. The effective area is 4108.47 deg$^{2}$} 
\label{fig:footprint}
\end{figure}

The \sample footprint is constructed directly from the high resolution \healpix \cite{2005ApJ...622..759G} maps given in \cite{y3-gold}. The main requirements imposed in the footprint are: Each pixel has to be observed at least once in $griz$, with a coverage greater than $80\%$. Pixels affected by foreground sources like extended galaxies or bright stars and regions affected by image artifacts are removed. Pixels with 10-$\sigma$ limited depth of $i_{\var{SOF},{\rm lim}}<22.3$ are removed, consistent with the faintest magnitudes in the \sample, see Eqs.(\ref{eq:magcut},\ref{eq:zrange}).  Pixels with a 10-$\sigma$ magnitude limit in $r_\var{SOF}$ and $z_\var{SOF}$ bands such that  $2\,r_{\var{SOF},{\rm lim}} -  z_{\var{SOF},{\rm lim}} < 24$ are also removed,
in order to ensure reliable measurements of the color defined in Eq.~(\ref{eq:colorcut}). The area of the \sample covers the 4108.47 deg$^{2}$ and is shown in Fig.~\ref{fig:footprint} as a projected density field. 

\subsection{Observational Systematics}

The observed number of galaxies is expected to have a non-trivial selection function that depends on various observing conditions of the survey, in addition to external conditions such as dust extinction or the dependence of stellar density with galactic latitude. These properties, which are themselves correlated, vary spatially and in general will have large-scale modes. This imprints a bias in the clustering signal if not accounted for \cite{ross11,Leistedt14,Ross:2017,Laurent:2017,Bautista:2018,Icaza-Lizaola:2020,Vakili:2020,Weaverdyck_2021}. We correct this effect by applying
weights to each galaxy corresponding to the inverse of the estimated angular selection function. This methodology, while now more widely adopted in the literature, was first applied in DES for the lens sample used in the DES Y1 3x2pt analysis in \cite{elvinpoole}. In Y3, the same methodology is applied to all clustering samples. Details about the method and results for other Y3 samples are given in \cite{y3-galaxyclustering} and for the \sample in \cite{y3-baosample}.

The method consists on assigning weights to correct for the spurious clustering signal from individual survey properties, iteratively in order of decreasing significance until a global threshold is met. In the  \gold catalogue, more than 100 survey property maps are available. However the majority of these maps are highly correlated among themselves. First we use a criterion to eliminate the highest correlations based on their Pearson's correlation coefficient matrix. This reduces the list of maps to a subset of 26 maps, including depth, airmass, stellar density and E(B-V) extinction (see Appendix B of \cite{y3-baosample}). We start by measuring the galaxy density as a function of survey property, for each tomographic bin separately. We use a set of 1000 lognormal realisations to measure what this relation is expected to look like in the absence of any induced systematic, and hence to estimate the significance of the relation found on the data. We then assign position-dependent weights to galaxies to remove the most significant trend in question. All weights are assigned with a linear fit to the density-observing condition relation, and we find no evidence for requiring additional terms in the model. This process is run iteratively until all density-observing conditions relations are below a given threshold. In the case of the \sample, we chose the threshold to be equivalent to $<99\%$ of mock values. 

Contrary to Y1, in Y3 we find that the observational systematic correction is several times the statistical error, mostly due to the increase in sample size. During the blinded period we tested variations in the treatment of systematics, e.g. masking regions with extreme survey property values or varying the significance threshold. They all led to consistent final weights maps. Moreover by comparison to a set of 1000 lognormal mocks, we find that the systematic error on the correction from over-fitting was small compared with the statistical errors in the angular correlation function or angular power spectrum itself. In Sec.~\ref{sec:robustness_tests}  we show that the recovered BAO distance measurement is insensitive of the observational weights.
More details about the mitigation of systematics can be found in \cite{y3-baosample}.

\section{Simulations}
\label{sec:simulations}

In what follows, we discuss the two sets of mock galaxy catalogues that we use throughout the analysis to validate the whole BAO distance measurement.
 
\subsection{ICE-COLA (quasi-n-body) Mocks}

We \red{create a set of 1952 mock catalogues} of the DES Y3 BAO sample that reproduce with high accuracy the principal properties of the data: (i) the sample observational volume, (ii) the abundance of galaxies, true redshift distribution and photometric redshift uncertainty and (iii) the clustering as a function of redshift. We refer the reader to \cite{y3-baomocks} for further details, and highlight here only the basic features of the ICE-COLA mocks used for this work.

This set of mocks is obtained using $N_{\rm sim}= 488$ fast quasi-$N$-body simulations generated with the ICE-COLA code \citep*{ice-cola}. The COLA method \citep{2013JCAP...06..036T,cola} uses second order Lagrangian Perturbation Theory (2LPT) combined with a Particle-Mesh (PM) gravity solver. The latter is used to solve particle trajectories on small scales, where the 2LPT accuracy is lower compared to the full $N$-body solution. The ICE-COLA algorithm extends the COLA algorithm to produce on-the-fly light-cone halo catalogues and weak lensing maps. The simulations use $2048^3$ particles in a box of size of $1536$~Mpc~$h^{-1}$, matching the mass resolution and 1/8 of the volume of the MICE \textit{Grand Challenge} simulation \citep{MICE1,MICE2}. A total of 64 box replications (four boxes in each Cartesian direction) are needed to create a full sky light-cone up to redshift $\sim 1.43$.

Mocks of galaxies are created populating halos following the recipe of a hybrid Halo Occupation Distribution and Halo Abundance Matching model \cite{2015MNRAS.447..646C,2018MNRAS.479...94A}. The algorithm has two free parameters per tomographic bin, setting the satellite number and the total number of galaxies as a function of host halo mass. These total  of ten free parameters are found by running an automatic likelihood minimization. This automatic calibration received as inputs the redshift distribution, $n(z)$,  and the unblinded data measurements of angular clustering  $w(\theta)$ at scales smaller than 1 deg. Another important property that needs to be covered by the mocks is a  realistic photometric redshift distribution. In contrast to the data, for the simulated halos we have the true redshifts and the photometric ones need to be derived. This is done by using the 2D probability  distribution  $P (z_{\rm ph},z_{\rm sp})$ of galaxies present in both VIPERS and DES Y3 data sets. Finally, four  non-overlapping DES Y3 footprint masks are placed on each full-sky halo catalogue allowing to have 4 times more galaxy mocks  than the total number of full-sky simulations.

The replications mentioned above introduce strong correlations among the measured $w(\theta)$ of tomographic bins that are not adjacent. This is discussed more in detail in  \cite{y3-baomocks} where it is shown that up to a $\sim10$\% of the particles are repeated (depending on the tomographic bin combination) once we impose the DES Y3 footprint and $n(z)$. This leads to non-zero covariances for tomographic bins that have no redshift overlap otherwise. For this reason we avoid using the ICE-COLA mocks as one of our primary covariance estimation tools, but use them only to validate and benchmark the process of angular diameter distance measurements. 

\subsection{FLASK (lognormal) Mocks}

Lognormal distributions have been shown to be a very good approximation to cosmological fields (\cite{Coles:1991if, Clerkin:2016kyr}). 
For some applications, they are a very useful tool due to their flexibility and for being much less computationally expensive than full N-body simulation runs.

We produce a set of lognormal mock catalogs using the publicly available code Full sky Lognormal Astro fields Simulation Kit (FLASK) \cite{Xavier:2016elr}. FLASK is able to quickly generate random catalog realisations, in tomographic bins, with the same statistical properties of the DES Y3 BAO galaxy sample. 

We generate 2000 mock galaxy position catalogs and density maps with the DES Y3 footprint in \healpix resolution NSIDE = 4096. We use as simulation input the galaxy bias and number density per tomographic bin, the full set of auto and cross correlations, and the lognormal field shift parameters\footnote{An additional parameter to specify the minimum value of the distribution.} per bin, as defined in \cite{Xavier:2016elr,y3-covariances}. The set of input correlations are the mean values measured in the ICE-COLA mocks. The galaxy biases used in the FLASK simulations are, namely, $b = [1.576, 1.595, 1.694, 1.821, 2.033]$ for the respective five redshift bins (also extracted from the ICE-COLA mocks). The cosmology adopted for FLASK is the {\tt MICE} cosmology. Considering this cosmology and the $n(z)$ of sample, the lognormal shift parameters for the five tomographic bins are  [0.600, 0.595, 0.593, 0.572, 0.580], respectively. 

For the angular power spectrum estimates, we convert the FLASK catalogs to \healpix maps with NSIDE=1024, and then measure the auto and cross spectra using the NaMaster code\footnote{\url{https://github.com/LSSTDESC/NaMaster}} \cite{2019MNRAS.484.4127A}, with the same specifications as with the real data (see Sec.~\ref{sec:measurements}). In real space, the auto and cross correlations of each of the 2000 catalogs were measured using TreeCorr\footnote{\url{https://rmjarvis.github.io/TreeCorr}} \cite{2004MNRAS.352..338J}. We set the \texttt{bin\_slop} parameter as $bs=0.0$ which means a brute force computation of the 2-point estimators.

We use these measurements to validate our baseline covariance based on the code \cosmolike \cite{cosmolike} (see Sec.~\ref{sec:covariance}).

\section{Analysis}
\label{sec:analysis}

\subsection{Clustering Measurements}
\label{sec:measurements}

We measure the clustering signal on the \sample using three different statistics, the angular correlation function $w(\theta)$ (ACF), 
the angular power spectrum in spherical harmonics, $C_\ell$ (APS) and the three-dimensional correlation in terms of projected comoving separation $\xi_p(s_\perp)$.

\subsubsection{Angular correlation function: $w(\theta)$}

The angular correlation function is computed after creating a uniform random sample within the mask defined in Sec.~(\ref{sec:mask}) with a size of 20 times that of the data sample in each tomographic bin. 
As pointed out in Sec. \ref{sec:mask} the mask has a pixel resolution of 4096 but includes a fractional coverage per pixel. We downsample the randoms according to this coverage and keep only pixels with coverage equal or larger than $80\%$. Given the random sample, we use the well known Landy-Szalay estimator \cite{LS1993}
\begin{equation}
w(\theta) = \frac{DD(\theta)-2DR(\theta)+RR(\theta)}{RR(\theta)},
\end{equation}
where \red{$DD$, $DR$ and $RR$ being the normalised counts of data-data, data-random and random-random pairs, with angular separation
$\theta\pm\Delta\theta/2$, with $\Delta\theta$ being the bin size, and all pair-counts are normalized based on the total size of each sample}. We bin pair-counts at a bin size of 0.05 degrees but later combined these into larger bin sizes to explore the dependence of the BAO fit statistics on angular resolution or size of the covariance matrix. As we will see (e.g. Figure~\ref{fig:ACF-data}), the BAO feature appears at $\sim$2.5 to $\sim$3.5 degrees at the redshift range considered here and has a width of approximately 1 degree. Hence, any coarse-graining due to this primary binning is not expected to affect the BAO measurements. We compute the clustering using two different pair counting codes, TreeCorr \cite{2004MNRAS.352..338J} and CUTE \cite{cute}, performing an extensive code-comparison to ensure consistent results. Finding excellent agreement between codes, we use CUTE with the brute force configuration as the default  clustering code for the rest of the analysis. After the tests in section \ref{sec:mocktest} we adopted a fiducial bin-size of $\Delta \theta = 0.2 \deg$, $\theta_{\rm min}=0.5 \deg$ and $\theta_{\rm max}=5 \deg$, which yields $N=22$ angular bins in total.\\

\subsubsection{Angular power spectrum: $C_\ell$}
\label{sec:cls}

We measure the angular power spectrum of galaxies using the so-called ``Pseudo-$C_\ell$'' (PCL) estimator \citep{2002ApJ...567....2H} as implemented in the \namaster code \citep{2019MNRAS.484.4127A}.
For constructing galaxy overdensity maps, we use the \healpix equal-area pixelization scheme, with a resolution of ${\rm NSIDE} = 1024$, corresponding to a mean spacing of $\sim 0.06$ degrees.
The equal-area pixelization allows computing the galaxy overdensity as $\delta_g = (N_p / w_p) (\sum_p w_p/ \sum_p N_p) - 1$ where $N_p$ is the number of galaxies in pixel $p$ \red{and $w_p$ the pixelized mask, that gives the fraction of the area of pixel $p$ covered by the survey}.

The discrete nature of galaxy number counts introduces a noise contribution to the estimated power spectrum, also known as noise-bias.
We assume this noise to be Poissonian, estimate it analytically following \citep{2019MNRAS.484.4127A,2020JCAP...03..044N}, and subtract it from our power spectrum estimates. Deviations from the Poissonian approximation are expected to be captured by broad-band terms in our template.

We bin the angular power spectrum estimates into bandpowers, assuming equal weight for all modes.
We use piecewise-linear, contiguous bins starting at a minimal multipole of $\ell_{\rm min}=10$ up to $\ell=1000$ and different values of $\Delta\ell$, chosen to guarantee a good signal-to-noise ratio across the bandpowers and remain flexible for scale cuts. After the tests in section \ref{sec:mocktest}, we adopted as fiducial choices $\ell_{\rm min}=10$, $\Delta\ell=20$ and an $\ell_{\rm max}$ scale-cut approximately corresponding to a $k_{\rm max}=0.25 h\, {\rm Mpc}^{-1}$ under the Limber relation, $k_{\rm max} = \ell_{\rm max} / r(\bar{z})$, evaluated at the mean redshift of each tomographic bin and the fiducial cosmology of our analysis. This $\ell$-binning allows us to resolve a BAO cycle with approximately 7 points (see \autoref{fig:APS-data}). 
When constructing the likelihood, we consider the effect of bandpower binning on the theory predictions using bandpower windows that account, in that order, for the effect of mode-coupling, binning averaging, and decoupling, following Sec 2.1.3 of \cite{2019MNRAS.484.4127A}.

\subsubsection{Projected Clustering: $\xi_{p}(s_\perp)$}
\label{subsec:projectedclustering}

In spectroscopic surveys, it is a common practice to transform redshifts to distances in order to measure physical comoving distances between galaxies. This enables access to three-dimensional information and measurements of the BAO shift parameter along and across the line-of-sight. 

\cite{Ross:2017emc} showed that for photometric surveys, whereas the radial clustering signal is erased due to the redshift uncertainty, angular BAO information \red{remains intact albeit smeared in the radial direction}. That paper also showed that for a DES-like BAO sample \red{representing the two-point correlation function as a function of the apparent perpendicular comoving distance ($s_\perp$), for different orientations with respect to the line-of-sight ($\mu$), leads to a BAO position that aligns very well as a function of $\mu$}. Hence, following \cite{Ross:2017emc}, we measure the anisotropic clustering $\xi(s_\perp,s_\parallel)$ using the Landy-Szalay estimator \cite{LS1993}:

\begin{equation}
    \xi(s_{\perp},s_\parallel) = \frac{DD(s_{\perp},s_\parallel)-2\cdot DR(s_{\perp},s_\parallel) +RR(s_{\perp},s_\parallel)}{RR(s_{\perp},s_\parallel)}
\end{equation}
with the normalised pair-counts separated by $s_{\parallel}$ and $s_{\perp}$ along and across the line-of-sight, respectively. We remark that distances $s_{\perp}$ and $s_{\parallel}$ are obtained by transforming photometric redshift and angular positions to comoving positions using the {\tt MICE} fiducial cosmology. Given the aforementioned alignment of BAO, we can combine our measurements into 
\begin{equation}
\xi_p({s_{\perp})=\int_0^{1} w(\mu) \xi\big(s_\perp(s,\mu),s_\parallel(s,\mu)\big)\ d\mu} \, ,
\label{eq:3D}
\end{equation}
\red{where $w(\mu)$ would \textit{a priori} be an inverse variance weighting. Ref.~\cite{Ross:2017emc} found that for $\mu < 0.8$ and $\sigma_z \ge 0.02 (1+z)$, our typical photo-$z$ error, the BAO signal is at a nearly constant $s_{\perp}$ while the signal for $\mu > 0.8$ is greatly diminished. Hence, for simplicity, one can approximate $w$ by} 
\begin{equation}
 w(\mu) = \begin{cases} 
      \frac{1}{0.8} & \mu\leq 0.8 \\
      0 &  \mu > 0.8. \\
   \end{cases} 
\end{equation}

Additionally, we could add a redshift-dependent FKP-like (\citep{FKP}) weight including the redshift uncertainty (Eq.~(16) of \cite{Ross:2017emc}). This per-galaxy weight is used to account for the change of BAO signal-to-noise ratio with redshift. We estimated it to be relatively flat, and we neglect it for the measurements shown here.  

This estimator has the advantage that all the galaxies in the full redshift range $0.6<z_{\rm ph}<1.1$ can be combined into a single clustering measurement with the full accumulated BAO signal, as opposed to splitting them into redshift bins. \cite{Ross:2017emc} showed that this estimator could reduce the statistical uncertainty with respect to using the angular correlation function. However, we found in \cite{2019MNRAS.483.4866A} that the modelling was slightly less robust and argued that this could be due to the Gaussian assumption for the redshift uncertainties. For this reason this was not included in the fiducial analysis for Y1 and for this work. We only include it in this study for visual purposes. 
After the completion of this work, we have submitted a new study to improve the modelling of $\xi_p$ \cite{Chan21} that may be applied in follow-up works to the DES data.

For this estimator, we perform pair-counts with $\Delta s_\perp=5 {\rm Mpc}/h$, $\Delta s_\parallel=1 {\rm Mpc}/h$, $s_{\perp,{\rm max}}=175 {\rm Mpc}/h$, $s_{\parallel,{\rm max}}=120 {\rm Mpc}/h$. 
With the fine binning in $s_\parallel$ being necessary to integrate $\mu$ in \autoref{eq:3D} and the maximum sizes set to avoid excessive computing resources.
The $\Delta s_\perp=5 {\rm Mpc}/h $ choice is relatively standard for $\xi(r)$ analysis.
Nevertheless, we remark that no BAO fits are derived in this paper from this estimator and we refer to the follow-up work on this estimator for further details and analysis choices \cite{Chan21}.

\subsection{BAO Template}
\label{sec:template}

We extract the BAO distance measurement from the clustering signal using a template-based method. This approach has been extensively used in the literature, mostly for spectroscopic datasets but also for photometric ones (e.g. \cite{Chan:2018gtc}). The main difference for the latter case is that one can extract mainly the angular diameter distance.

The main difference with previous template based BAO analysis, and in particular our DES Y1 results, is that our template is now fully obtained from first principles, including the damping of the BAO features. We implement this by means of the resummation of infrared (long wavelength) modes put forward in \cite{2015JCAP...02..013S,2016JCAP...07..028B} and others. This replaces the previous methodology of calibrating the BAO damping to be used in the data with mock simulations, and enable us to easily change the template for different cosmologies. We build a template for both the configuration space and harmonic space analysis. 

We build the BAO template starting from a linear power spectrum $P_{\rm lin}(k)$ obtained with {\sc Camb} \citep{2000ApJ...538..473L}. At BAO scales the main modification 
due to non-linear evolution is the broadening of the BAO feature due to large-scale flows.  We model this by introducing a Gaussian damping of the BAO wiggles, after isolating this component from the full power spectrum shape: 
\begin{equation}
P(k,\mu) = (b+\mu^2 f)^2\left[ (P_{\rm lin}-P_{\rm nw})e^{-k^2\Sigma_{\rm tot}^2}+P_{\rm nw}\right],
\label{eq:pkmu}
\end{equation}
where $P_{\rm nw}$ describes the smooth shape of the power spectrum and all the BAO information is in $P_{\rm lin} - P_{\rm nw}$.  In Eq.~(\ref{eq:pkmu}), $\mu \equiv {\rm cos}(\theta_{\rm LOS})$ = $k_{||}/k$, $b$ is the linear galaxy bias and  $f$ is the logarithmic derivative of the growth factor with respect to the scale factor (at the given cosmology) evaluated at the effective redshift of the sample. The pre-factor in Eq.~(\ref{eq:pkmu}) accounts for linear-theory redshift space distortions \citep{1987MNRAS.227....1K}. 
The bias parameters $b$ are obtained by fitting to $w(\theta)$ measurement from the data at three linear bins between 0.5 and 1 degree \cite[see details in][]{y3-baomocks}. Note that these scales do not contain any BAO information, hence these measurements do not interfere with the blinding scheme. We also note that the non-linearities at small scales are expected to only appear below 0.3 deg., see \cite{Krause21}. 

There are several methods to define the smooth ``no-wiggle'' power spectrum, $P_{\rm nw}(k)$. We follow the 1D Gaussian smoothing in log-space described in Appendix A of \cite{2016JCAP...03..057V}. We start by
defining the ratio of $P$ and a smooth approximation to it, for which we employed the no-wiggle fitting formulae from \cite{1998ApJ...496..605E} (hereafter EH). This reduces the dynamical range and makes the
filtering more efficient. Then
\be
P_{\rm nw} (k) = P_{\rm EH}(k) \times \left[ {\mathcal F} * R \right](k)
\label{eq:pknw}
\ee
where $R(k) = P(k) / P_{\rm EH}$ and the convolution with the filter ${\mathcal F}$ is in log$_{10}$ variables,
\be
\left[ {\mathcal F} * R \right](k) = \frac{1}{\sqrt{2\pi} \lambda} \int d\log q R(q) \exp\left[ -(\log (k / q))^2 / 2 \lambda^2\right] \nonumber.
\ee
We use $\lambda = 0.25$, a standard value for which we recover $P_{\rm nw} \rightarrow P_{\rm lin}$ at low and large $k$.

For the resummation of infrared modes, leading to the Gaussian damping of the BAO feature in Eq.~(\ref{eq:pkmu}), we take as a reference the implementation in \cite{2020JCAP...05..042I,2018JCAP...07..053I} and write, 
\be
\label{eq:Sigma_tot} 
\Sigma^2_{\rm tot} (\mu) = \mu^2 \Sigma_{\parallel}^2 + (1-\mu^2)\Sigma_{\perp}^2+f\mu^2(\mu^2-1)\delta\Sigma^2 
\ee
where $\Sigma_{\parallel}=(1+f) \Sigma$ and $\Sigma_{\perp}=\Sigma $, with
\bea
\Sigma^2&=&\frac{1}{6\pi^2}\int_0^{k_s} dq \, P_{\rm nw}(q) \left[ 1- j_0(q \ell_{\rm BAO}) + 2 j_2(q \ell_{\rm BAO})\right] \, \nonumber \\
\delta \Sigma^2&=&\frac{1}{2\pi^2}\int_0^{k_s} dq \,P_{\rm nw}(q) j_2(q \ell_{\rm BAO}) 
\eea
where $j_n$ are the spherical Bessel functions of order $n$, while $\ell_{\rm BAO}$ is the correlation length of BAO. As a reference we choose values $k_s=0.2 \invmpc$ and $\ell_{\rm BAO} = 110 \mpc$, but our results do not depend on these choices. We assume a damping that scales with the growth factor: $\Sigma = \Sigma_0 \cdot D(z)$ and  $\delta \Sigma = \delta \Sigma_0 \cdot D(z)$. For the {\tt MICE} cosmology we obtain $\Sigma_0 = 5.80\mpc$ and $\delta \Sigma_0 = 3.18\mpc$ while for the {\tt Planck} cosmology  we find $\Sigma_0 = 5.30\mpc$ and $\delta \Sigma_0=2.81\mpc$. 

As in \cite{Chan:2018gtc},  we have also determined $\Sigma$  (i.e. $\Sigma_{\rm tot } $ from Eq.~(\ref{eq:Sigma_tot}) without the $\delta \Sigma$ term) directly by fitting to the mean of the COLA mocks using different values of $ \Sigma $.  The best fit (minimum $\chi^2$) $\Sigma $ is $ 5.85 \mpc$ and it is fully consistent with the analytical {\tt MICE} cosmology result from Eq.~(\ref{eq:Sigma_tot}).  

Once provided with $P(k,\mu)$ we compute the anisotropic redshift-space correlation function $\xi(s,\mu)$ through a Fourier transform. The angular correlation function is obtained after projecting $\xi$ weighted by the redshift distribution $n(z)$  (normalised to 1),
\be
\label{eq:w_template_schematic}
w(\theta) = \int dz_1\int dz_2 n(z_1) n(z_2)\xi \big( s(z_1,z_2,\theta),\mu(z_1,z_2,\theta) \big).
\ee
To compute the $C_\ell$ template, we first evaluate $w(\theta)$ from $0.001\deg$ to $179.5 \deg$, in 300 steps with logarithmic spacing and then transform it to $C_\ell$ by integrating numerically,
\be
\label{eq:Cl_template_Ltransform}
C_\ell = 2 \pi \int_{-1}^{1} d(\cos \theta) \,  w(\theta)  P_\ell (\cos \theta)
\ee
where $P_\ell$ is the Legrendre polynomial of order $\ell$.  In this way the two baseline templates are strictly consistent with each other\footnote{Besides, the $ \xi_{\rm p}$ template is also obtained by rebinning of $w$ (see \cite{Chan21}); thus, the $\xi_{\rm p}$ template is also consistent with the others.}. 

The template is finally composed by the piece containing BAO information described above and  a set of terms without BAO information,
\be
\label{eq:template_all_parameters}
M(x) = B T_{\rm BAO, \alpha}(x^\prime) + A(x), 
\ee
where we include the parameter $B$ to allow adjustment of the overall amplitude and the function $A$ is a smooth function designated to absorb the imperfectness of the full shape template modelling and the remaining systematic contributions.   

In the case of $w$, $x=\theta$, $T$ corresponds to $w$ as given by Eq.~(\ref{eq:w_template_schematic}), $x^\prime=\alpha \theta $, where $\alpha $ is the BAO shift parameter we are after, and the function $A$ takes the form
\be
\label{eq:broadband_terms_w}
A(\theta)  = \sum_i \frac{ a_i }{ \theta^i }.
\ee
For $C_\ell$,  $x=\ell$, $T$ is computed via Eq.~(\ref{eq:Cl_template_Ltransform}), $x^\prime= \ell / \alpha $, and  $A$ is of the form
\be
\label{eq:broadband_terms_Cl}
A(\ell)  = \sum_i  a_i  \ell^i . 
\ee
We will consider different ranges for the $i$ index in Sec.~\ref{sec:mocktest}, see Tab. ~\ref{tab:Cl_mocktest}.

\subsection{Covariance Matrix}
\label{sec:covariance}

\begin{figure*}
	\centering
	\includegraphics[width=\textwidth]{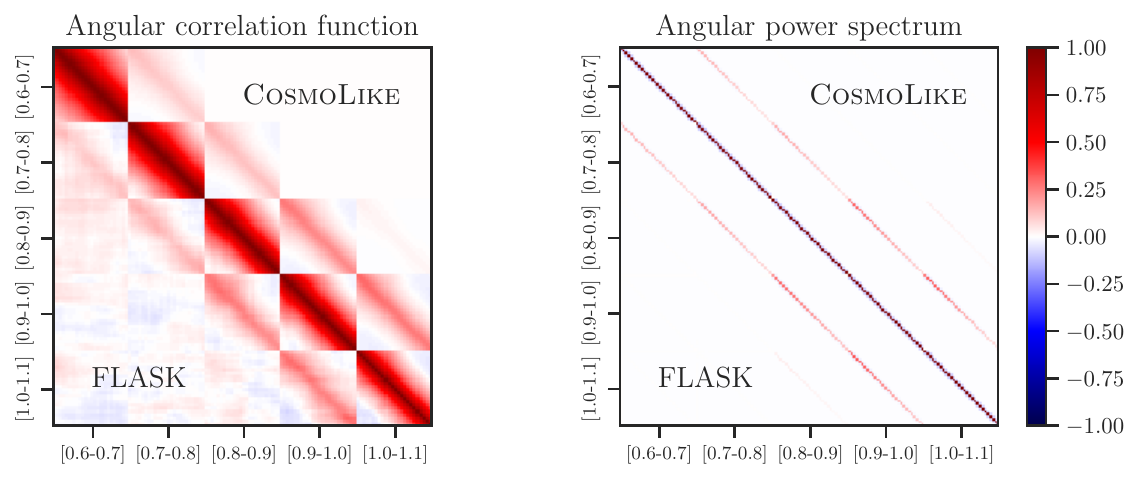}
	\caption{Comparison of correlation matrices obtained from \cosmolike (upper right triangles) and from 2000 FLASK realizations (lower left triangles). The left panel shows the ACF correlation matrix, while the right panel corresponds to the APS one.    }
	\label{fig:cov_comparison}
\end{figure*}

We rely on analytic estimates for our fiducial covariance computation, which we validate against covariances estimated from mocks. Following \cite{Crocce2011}, the real space covariance of the angular correlation function $w(\theta)$ at angles $\theta$ and $\theta'$ is related to the covariance of the angular power spectrum by
\begin{equation}
\begin{aligned}
&\mathrm{Cov}(w(\theta),\, w(\theta')) = \\
 &\sum_{\ell, \, \ell'}\dfrac{(2\ell +1)(2\ell'+1)}{(4\pi)^2}\overline{P_{\ell}}(\theta)\overline{P_{\ell'}}(\theta')  \mathrm{Cov}(C_{\ell}, C_{\ell'}) ,
 \label{eq:Cov_wtheta}
\end{aligned}
\end{equation}
where $\overline{P_{\ell}}(\theta)$ are the Legendre polynomials averaged over each angular bin $[\theta_{\min}, \, \theta_{\max}]$ and are defined by
\begin{equation}
\overline{P_{\ell}} = \dfrac{\int_{x_{\min}}^{x_{\max}} \der x \, P_{\ell}(x) }{x_{\max} - x_{\min}} = \dfrac{\left[ P_{\ell+1}(x) - P_{\ell - 1}(x) \right]_{x_{\min}}^{x_{\max} }}{(2\ell+1)(x_{\max} - x_{\min}) },
\end{equation}
with $x= \cos \theta$ and $x_{\{\min, \, \max\}} = \cos \theta_{\{\min, \max\}}$. 
In principle, the indices in \autoref{eq:Cov_wtheta} run for all $\ell$s individually from 0 to $\infty$, although we stop at a large $\ell_{\rm max}$, once convergence is reached.

The covariance matrix $\mathrm{Cov}(C_{\ell}, C_{\ell'})$ can be split into a Gaussian term (that does not include higher-order moments of the density field), and a non-Gaussian term that involves the 4-point function of the density field (the trispectrum) \cite{Takada2009} and a super-sample covariance contribution \cite{Takada2013}. We have tested that including these does not impact our results, therefore our fiducial covariance only includes the Gaussian terms. In that case, the covariance of the angular power spectrum in a given tomographic bin is given by \cite{Crocce2011,cosmolike}
\begin{equation}
\mathrm{Cov}(C_{\ell}, C_{\ell'}) =  \dfrac{2\delta_{\ell \ell'}}{f_{\rm sky}(2\ell + 1)}\left(C_{\ell'} + \frac{1}{n_g}\right)^2,
\label{eq:cov-c_ell}
\end{equation}
where $\delta$ is the Kronecker delta function, $n_g$ is the number density of galaxies per steradian, and $f_{\rm sky}$ is the observed sky fraction,  which is used to account for partial-sky surveys.

We use the \cosmolike code to compute the analytical covariance matrices \cite{cosmolike,cosmolike2020,cosmolike_curvedsky}. We include redshift space distortions through the $C_{\ell}$'s of Eq.~\ref{eq:cov-c_ell} and, following \cite{cosmolike_mask}, we correct the shot-noise contribution to the covariance (the term $\propto1/n_g$) by taking into account the effect of the survey geometry to the number of galaxies in each angular bin $\Delta \theta$. 

For the harmonic space analysis, we begin with the \cosmolike predictions for the angular power spectra and compute analytical Gaussian covariance matrices accounting for broadband binning and partial sky coverage in the PCL estimator context, following \cite{2004MNRAS.349..603E,2019JCAP...11..043G}. We compute the coupling terms using the \namaster implementation \cite{2019JCAP...11..043G,2019MNRAS.484.4127A}.

We have validated both our real space and harmonic space analytic covariance matrices with estimations from simulations, which are described in Section~\ref{sec:simulations}. The predicted $w(\theta)$ and $C_{\ell}$ from \cosmolike are in very good agreement with the measurements from the mocks, and we also obtain consistent results when using a covariance estimated from either COLA or FLASK mocks (see Tables~\ref{tab:w_mocktest} and \ref{tab:Cl_mocktest}). In Figure~\ref{fig:cov_comparison} we show that there is good agreement between the correlation matrices obtained from \cosmolike and FLASK mocks. The covariances estimated from COLA \red{are not shown in the Figure~\ref{fig:cov_comparison} and} have larger cross-covariance elements due to a replication problem with the mocks, as explained in \cite{y3-baomocks}. We refer the reader to \cite{y3-baomocks} for a comparison between \cosmolike and COLA covariances.

\subsection{Parameter Inference}
\label{sec:inference}

The likelihood function of the parameters $\bm{p}$ given the data  $ \bm{d}$, $\mathcal{L}$,  measures the goodness of the model fit to the data. Under the Gaussian likelihood approximation, the likelihood is related to the $\chi^2 $ as 
\begin{align}
\mathcal{L} ( \bm{p} |  \bm{d}  ) \propto e^{ - \frac{\chi^2}{2} } ,
\end{align}
where $\chi^2 $ is given by 
\begin{align}
\chi^2(\bm{p}  |  \bm{d}  ) =\sum_{ij} \big[ \bm{d} - \bm{M}( \bm{p} )\big]_i C^{-1}_{\, ij } \big[ \bm{d} - \bm{M}( \bm{p} )\big]_j,
\end{align}
with $C$ being the covariance matrix of $ \bm{d} $.

The best fit model $ \bm{M} $ can be estimated by looking for the parameters $\bm{p} $ at which the likelihood attains its maximum, i.e.~the maximum likelihood estimator.   The procedure of $\chi^2$ minimization given the nuisance parameters in Eq.~\ref{eq:template_all_parameters} are similar to those described in \cite{Chan:2018gtc}. We first analytically fit the linear parameters for the broadband terms $a_i $ in Eq.~\ref{eq:broadband_terms_w} or~\ref{eq:broadband_terms_Cl}. The residual $\chi^2$  is further numerically minimized with respect to the amplitude parameter $B$ subject to the condition that $B>0$. We are left with $\chi^2$ of single parameter $\alpha$, whose minimum gives the best fit $\alpha$.

In this work, ${\bf d}$ represents the data from the auto-correlation of each redshift bin. We studied in \cite{Chan:2018gtc} the possibility of including cross-correlations between bins, finding that the gain was very small and at the cost of increasing significantly the size of the data vector and its covariance.
 
In the case of Gaussian likelihood,  the 1-$\sigma$ error bar for one single parameter ($\alpha$ in our case) is given by the condition that  $\Delta \chi^2 \equiv  \chi^2(p) - \chi^2(p_0)  =1 $, where $\chi^2( p_0 )$ is evaluated at the best fit $p_0$. We will quote the 1-$\sigma$ error bar derived from this criterion.  We will see in Sec.~\ref{sec:mocktest} that this error bar agrees reasonably well with the distribution of the best fit $\alpha$ from the mock results. However, there remains small but non-negligible differences, which indicate deviation from the Gaussian likelihood; thus, we will provide the likelihood for $\alpha$ when the cosmological constraints are desired. We use exclusively the frequentist $\chi^2 $ fitting to extract the best fit parameters, as we checked that it gives consistent results to the Bayesian method in \cite{Chan:2018gtc} (see also \cite{Cuceu_etal2020}).    

The template is computed in the fiducial cosmology, and the cosmological information is encoded in the angular BAO scale. The BAO shift parameter $\alpha $  bridges the angular BAO scales in the actual cosmology and the fiducial one as  
\be
\label{eq:alpha_BAOtransverse}
\alpha = \frac{D_M(z)}{r_{\rm d}} \frac{r^{\rm fid}_{\rm d}}{D^{\rm fid}_M(z)},
\ee
where the above expression is evaluated at the effective redshift of the sample, described below. In Eq.~(\ref{eq:alpha_BAOtransverse}) $r_{\rm d}$ is the sound horizon at the drag epoch, $D_M$ is the (comoving) angular diameter distance\footnote{We follow the definition of $D_M$ in Eqs. (15-17) of Ref \cite{2021PhRvD.103h3533A}.} and ${\rm fid}$ denotes the fiducial cosmology used for the analysis. 

We define the effective redshift as the weighted mean redshift of the sample
\begin{equation}
\label{eq:zeff}
    z_{\rm eff} = \frac{ \sum_i w_{i, {\rm sys}} \cdot w_{{\rm FKP}}(z_i) \cdot z_i} {\sum_i w_{i, {\rm sys}} \cdot w_{{\rm FKP}}(z_i)} = 0.835  \, ,
\end{equation}
with $w_{i, {\rm sys}}$ the systematic weight of the galaxies and the  $w_{{\rm FKP}}$ the statistical inverse-variance weight, see Eq. ~16 of \cite{Ross:2017emc}. This definition was also used for the DES Y1 BAO analysis. We note that alternative definitions can lead to changes in $z_{\rm eff}$ of up to $\Delta z_{\rm eff} \sim 0.035$. However, since in BAO measurements the $D_M(z_{\rm eff})$ is divided by the fiducial value  $D_M^{\rm fid}(z_{\rm eff})$, see Eq.~(\ref{eq:alpha_BAOtransverse}),  they are not very sensitive to changes in $z_{\rm eff}$, as long as we assume a smooth evolution of $D_M(z)$ both at the fiducial and underlying cosmology. For example, for a change from {\tt MICE} to {\tt Planck} cosmology, an error of $\Delta z_{\rm eff}= 0.035$ translates to an error of $\Delta \alpha = 0.001$, well below the statistical uncertainties reported here. 

In summary, for {\tt MICE} cosmology $r_{\rm d} = 153.4 \,  \mathrm{Mpc}$ while $D_M(0.835)=2959.7\, \mathrm{Mpc} $, leading to  $D_M/r_{\rm d} = 19.29$. For {\tt Planck} cosmology $r_{\rm d} = 147.6 \,  \mathrm{Mpc} $ while $D_M(0.835)=2967.01 \, \mathrm{Mpc} $, leading to $D_M/r_{\rm d} = 20.1$.

\subsection{Combining Statistics}
\label{sec:comb}

For consistency and robustness we derive angular distance measurements from real and harmonic space statistics. 
Even if the angular power spectrum (APS) and the angular correlation function (ACF) contain similar information, the two analyses have been developed using different techniques. In addition, the systematic uncertainties in each one could be different. The compatibility of the two
values is a robustness test of the measurement and, given their different sensitivities, both values  can be still combined to gain some precision 
in the final BAO scale measurement. However, the two measurements are highly correlated, and the combination must be done carefully, taking 
into account the high level of correlation. 

As we will see in Sec.~\ref{sec:mocktest} the standard deviations for both measurements are very similar, independently of the covariance matrix we used in the fits (we used COLA mocks, \cosmolike mocks and FLASK mocks). This means that the two analyses have similar sensitivity and similar statistical power, and that some margin for gains exists. To take into account these properties in our combination, we consider two different methods for combining the measurements.  \vspace{0.2cm}

$\bullet$ {\it Method 1} : In the first method we average the $\chi^2$ distributions for each measurement to obtain the combined result. This is equivalent to defining the combined likelihood as the geometric mean of the likelihoods from each space, $\mathcal{L(\alpha)_{\rm comb}} =  (\mathcal{L(\alpha)_{\rm APS}} \mathcal{L(\alpha)_{\rm ACF}})^{1/2}$. This approach is conservative in the sense that it ignores the correlation of the measurements and will tend to over-estimate the combined error. Nonetheless we will consider this our fiducial method. \vspace{0.2cm}

$\bullet$ {\it Method 2} : This assumes that the individual likelihoods are Gaussian and uses the combination of two correlated Gaussian variables to obtain the final result. Let us call $\rho$ the Pearson correlation
coefficient for the two sets of measurement. Then, the correlation matrix between the ACF and the APS $\alpha$'s is given by
\begin{equation}
    \text{corr}(\alpha^{\text{ACF}},\alpha^{\text{APS}})=\begin{pmatrix}1 &  \rho \\ \rho & 1\end{pmatrix}.
\end{equation}
Now, we define
\begin{equation}
    \sigma_1=\sigma_\alpha^{\text{ACF}} \quad \sigma_2=\sigma_\alpha^{\text{APS}}.
\end{equation}
Assuming correlated Gaussian distributions, the combined $\alpha$ and its error are given by
\begin{align}
  \alpha^{\text{combined}} & =w\alpha^{\text{ACF}}+(1-w)\alpha^{\text{APS}},\\
  \sigma_\alpha^{\text{combined}} &=\left[\frac{(1-\rho^2)\sigma_1^2\sigma_2^2}{\sigma_1^2+\sigma_2^2-2\rho\sigma_1\sigma_2}\right]^{1/2},  \label{eq:comb_error}
\end{align}
where $w$ is defined as
\begin{equation}
\label{eq:w_comb}
    w=\frac{\sigma_2^2-\rho\sigma_1\sigma_2}{\sigma_1^2+\sigma_2^2-2\rho\sigma_1\sigma_2}.
\end{equation}
Note that in those cases where the error difference is too large given the correlation coefficient the weights may be negative. In order to avoid negative contribution from one of the measurements, we will discard those cases.

As we will see below, the correlation coefficient is very high for the case of study here: $\rho=0.893$.

\begin{figure}
  \includegraphics[width=\linewidth]{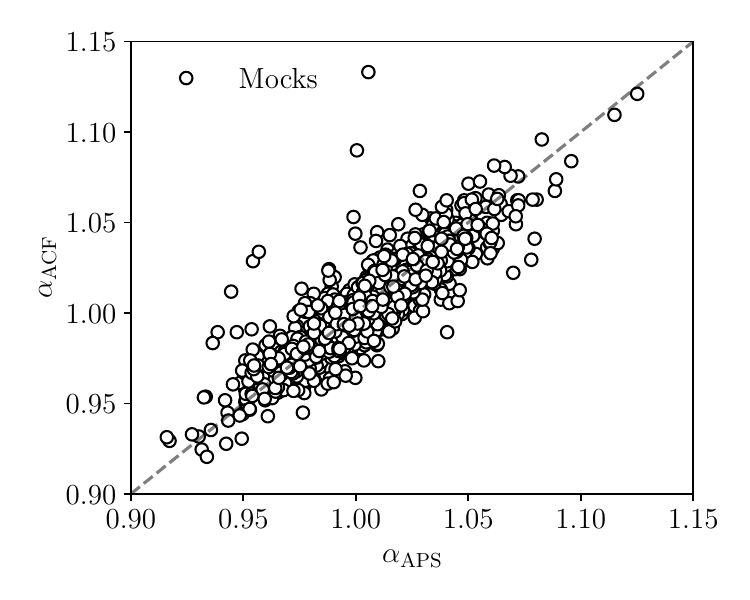}
  \includegraphics[width=\linewidth]{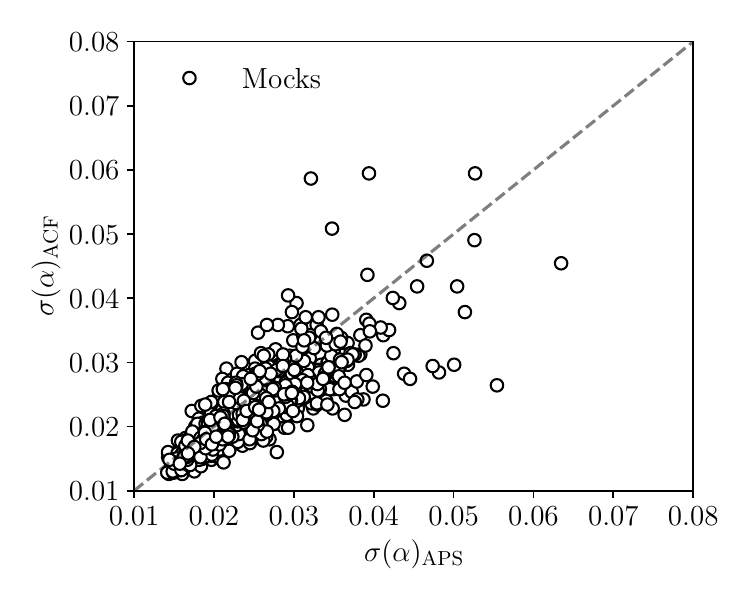}
  \caption{Comparison between the recovered BAO fit (top) and its uncertainty
    (bottom) from the ACF and the APS on the 1952 ICE-COLA mock realizations (white circles). The $1-\sigma$ errors from APS are $\sim 10\%$ larger than those from ACF, but agree better with the standard deviation from the $\alpha$ distributions on the top panel.}
  \label{fig:correlated_statistics}
\end{figure}

\section{Tests on Mocks}
\label{sec:mocktest}

We have carried out a series of tests to verify our BAO fitting pipeline using the 1952 COLA mocks.
The tests are similar to those performed in \cite{Chan:2018gtc,Camacho:2018mel,2019MNRAS.483.4866A}.  Table \ref{tab:w_mocktest} and \ref{tab:Cl_mocktest} show the major mock tests for  $w$  and  $C_\ell$ respectively. 

A number of metrics are used to quantify the accuracy of the BAO fitting procedures. Recall that in this paper we define the best-fit $\alpha$ through a maximum likelihood estimator and the 1-$\sigma$ error, $\sigma_\alpha$, through the condition $\Delta \chi^2 \equiv 1$, as discussed in Sec.~\ref{sec:inference}. Figure \ref{fig:correlated_statistics} shows the distribution of $\alpha$ (upper panel) and $\sigma_\alpha$ (lower panel) obtained in $w(\theta)$ vs. $C_\ell$ space.

In Table \ref{tab:w_mocktest} and \ref{tab:Cl_mocktest} the distribution of the best fit $\alpha$ derived from the mocks is quantified by its mean, $\langle \alpha \rangle $, and two measures of the spread of the distribution: the standard deviation $\sigma_{\rm std } $ and $\sigma_{ 68}$. The latter is defined as the symmetric error bar between the $16^{\rm th} $ and $84^{\rm th}$ percentile of the distribution and it is less sensitive to the tails of the distribution compared to $\sigma_{\rm std } $.  The mean of the error bar derived from the likelihood $\langle \sigma_\alpha \rangle $ is also shown.  For the error bar to be meaningful, it should agree with the measures of the spread of the distribution. Another way to quantify the accuracy of the error bar is to check the fraction of mocks enclosing $\langle \alpha \rangle $, i.e. with $ \langle \alpha \rangle -\sigma_\alpha \le \alpha \le \langle \alpha \rangle  + \sigma_\alpha$, which is 68\% from the Gaussian expectation.  The pull statistics $ d_{\rm norm} = ( \alpha -\langle \alpha \rangle ) / \sigma_\alpha $ enables us to study the correlation between the deviation of the individual best fit from the ensemble one and the error bar derived from the fit. We have shown the mean and the standard deviation of the distribution of $ d_{\rm norm} $, $\langle  d_{\rm norm} \rangle $ and $\sigma_{d_{\rm norm } }$.  The goodness of the fit is indicated by the mean chi-squared per degree of freedom,  $\langle \chi^2 \rangle/\mathrm{d.o.f.} $ To give a representative fit result, we show the fit to mean of mocks using the covariance for a single mock.
   
The fiducial cosmology for the mock test (template and covariance) is the {\tt MICE}
cosmology and the fiducial covariance estimation is from \cosmolike. For the
$w$ fit to the mocks, the fiducial template is computed with $ \Delta \theta =
0.2^\circ$, and the fitting is performed in the angular range
$\left[0.5^\circ,5^\circ\right]$.  The default number of broadband terms is
$\sum_i A_i / \theta^i$ with $i$ going over 0, 1, and 2.  For the APS, the
default setup is defined by linear piece-wise bandpowers of width
$\Delta\ell=20$ with $\ell_{\rm min}=10$ and $\ell_{\rm max}$ derived from a
sharp cut of $k_{\rm max} = 0.25\, h\, {\rm Mpc}^{-1}$ translated to each
tomographic bin using the Limber relation, yielding $\ell_{\rm max}=(410, 470,
510, 550, 610)$.  For the broadband model we used $A(\ell) = \sum_i A_i \ell^i$
with four terms for $i$ running from -1 to 2. The mock results for this
fiducial configuration are shown in bold fonts in Tables \ref{tab:w_mocktest}
and \ref{tab:Cl_mocktest}, where we also show the results from varying this
configuration in terms of the number of broad-band terms, min/max scales,
angular binning, and covariance estimation method.

The best fit $\alpha $ from the mean of the mocks is consistent with 1 with a small positive bias which can be attributed to nonlinear evolution of the BAO scale \citep{CrocceScoccimarro_2008,PadmanabhanWhite_2009}. This bias ($\sim 0.4\%$) is well below our expected statistical uncertainty ($\sim 2.2\%$) and therefore we consider it negligible and the template construction method described in Sec.~\ref{sec:template} as unbiased. 

In turn, we find that $ \langle  \sigma_\alpha \rangle  $ is slightly smaller than $\sigma_{\rm  std} $. The differences are smaller for $C_\ell$ than for $w$. On the other hand, $ \langle  \sigma_\alpha \rangle  $ is closer to $\sigma_{68} $. For $C_\ell$,  $ \langle  \sigma_\alpha \rangle  $ agrees with  $\sigma_{68} $ well, while for $w$,  $ \langle  \sigma_\alpha \rangle  $ is still slightly smaller than  $\sigma_{68} $. The discrepancy suggests deviations of the likelihoods from perfect Gaussianity, in particular the ACF. But these deviations are small, with the true errors being underestimated by $\sigma_\alpha$ by $\lesssim 10\%$ for ACF (and less so for APS). For both $w$ and $C_\ell$, these results are consistent with the fraction of mocks enclosing $\langle \alpha \rangle $ in each case, that for APS yields almost perfectly the expected $68\%$ and for ACF falls short by $\sim 10\%$. The distribution of $d_{\rm norm }$ is close to Gaussian distribution with zero mean and unit variance. The fact that  $\sigma_{ d_{\rm norm }}$  is slightly larger than 1 follows from the same trend as $ \langle  \sigma_\alpha \rangle $. 
Overall our model offers a good fit to the mock data and the error estimations are robust and consistent.  The ACF likelihood deviates slightly more from perfect Gaussianity than the APS but as we will see next these small discrepancies are removed once ACF and APS are combined.  

\begin{table*}
  \caption{BAO fits for the ACF using the \cosmolike covariance on the COLA mocks with different configurations and variations of the analysis.   The form of the broadband terms is $\sum_i A_i / \theta^i$. We also show the results from the angular binning $\Delta \theta$ and the minimum angular scale considered $\theta_{\rm min}$.  We have also tested the results obtained with  the COLA mock-based covariance or the one from FLASK mocks.  In the next row, {\tt Planck} cosmology is assumed for both the BAO template and the \cosmolike covariance. Finally, in order to check the number of broadband terms required to get an unbiased estimate for an alternative cosmology template, we have shown the results obtained with the {\tt Planck} cosmology template for different number of broadband terms.  The bold entries indicate our baseline choice, in addition to $\theta_{\rm min} = 0.5 \deg$ and $\theta_{\rm max}=5\deg$.}
  \label{tab:w_mocktest}
  \begin{tabular}{lccccccccc}
  \hline
  \hline
case &  $\langle \alpha \rangle$     &   $\sigma_{\rm std}$   &   $\sigma_{68}$      & $\langle \sigma_{\alpha} \rangle$    &  ${\rm fraction\,encl.}   \langle \alpha \rangle$   &  $\langle d_{\rm norm} \rangle$   & $\sigma_{d_{\rm norm}}$  &   $\langle \chi^2 \rangle / {\rm d.o.f.}$   & ${\rm mean\,of\,mocks}$        \T\B \\ \hline   
$i=0$               &  1.002   &  0.022  & 0.021  & 0.021  &  65\%  & 0.010   & 1.057   &  101.5/99 (1.03)   &  1.002 $\pm$ 0.020    \T \\  
$i=0,1$             &  1.003   &  0.024  & 0.023  & 0.022  &  63\%  & -0.025  & 1.090   &  97.6/94 (1.04)   & 1.003 $\pm$ 0.021        \\  
${\bf i=0,1,2}$ &  {\bf 1.004}   &  {\bf 0.024}  & {\bf 0.023}  & {\bf 0.021}  &  {\bf 62\%}  & {\bf -0.024}  & {\bf 1.116}   &  {\bf 93.1/89 (1.05)}   &  {\bf 1.004 $\pm$ 0.021}       \\  
$i=-1,0,1,2$        &  1.004   &  0.024  & 0.023  & 0.021  &  63\%  & -0.026  & 1.124   &  88.3/84 (1.05)   & 1.004 $\pm$ 0.021     \B \\  

$\Delta \theta = 0.10\deg$            & 1.004  &  0.024  & 0.023  & 0.021   & 64\%  & -0.019  & 1.107  &  198.2/204 (0.97)   &  1.004 $\pm$ 0.021  \T \\          
$\Delta \theta = 0.15\deg$            & 1.004  &  0.024  & 0.023  & 0.021   & 63\%  & -0.019  & 1.107  &  129.0/129 (1.01)   &  1.004 $\pm$ 0.021     \\          
$ {\bf \Delta \theta = 0.20\deg}$ & {\bf 1.004}  &  {\bf 0.024}  & {\bf 0.023} & {\bf 0.021}   & {\bf 62\%}  & {\bf -0.024} & {\bf 1.116}   &  {\bf 93.1/89 (1.05)}  &  {\bf 1.004 $\pm$ 0.021}  \B \\
  
$ \theta_{\rm min} = 1\deg$  &  1.004  &  0.024  &  0.023  &  0.021    &  62\%    & -0.028 & 1.120  &  82.8/79 (1.05)   &  1.004 $\pm$ 0.021  \T\B \\
  
$ {\rm COLA\,cov} $  & 1.004   &   0.025   &  0.024  &  0.023    & 66\%    & -0.024 & 1.055   &  86.2/89 (0.97)   & 1.003$\pm$ 0.023 \T  \\ 
$ {\rm FLASK\,cov} $ & 1.004   &   0.026   &  0.025  &  0.022    & 64\%    & -0.023 & 1.098   &  90.4/89 (1.02)   & 1.003$\pm$ 0.022  \B \\

$ {\rm {\tt Planck} \,cosmology}$ & 0.966 &   0.023   &  0.023  &  0.026    & 73\%    & -0.017 & 0.880   &  72.7/89 (0.82)   & 0.965$\pm$ 0.026  \B \\

$ {\rm {\tt Planck} \,temp.} \, i=0 $        & 0.949 &   0.022   &  0.022  &  0.021    & 64\%    & 0.048  & 1.05   &  109.9/99 (1.11)   & 0.949$\pm$ 0.022  \\
$ {\rm {\tt Planck} \,temp.} \, i=0, \, 1 $  & 0.965 &   0.023   &  0.023  &  0.024    & 69\%    & -0.019  & 0.98  &  101.5/94 (1.08)   & 0.965$\pm$ 0.023  \\
$ {\rm {\tt Planck} \,temp.} \, i=0, \, 1, \, 2 $  & 0.966 &   0.023   &  0.023  &  0.022    & 65\%    & -0.021  & 1.06  &  94.3/89 (1.06)   & 0.965$\pm$ 0.022  \\
$ {\rm {\tt Planck} \,temp.} \, i=-1, \, 0, \, 1, \, 2 $  & 0.966 &   0.024   &  0.022  &  0.022    & 65\%    & -0.027  & 1.07  &  89.0/84 (1.06)   & 0.966$\pm$ 0.022 \B \\

\hline
  \end{tabular}
\end{table*}

\begin{table*}
  \caption{BAO fits for the APS using the \cosmolike covariance on the COLA
    mocks with different configurations and variations of the analysis.  The
    form of the broadband terms is $A(\ell) = \sum_i A_i \ell^i$.  We also show
    the results from the broadband binning $\Delta \ell$ and the maximum
    multipole considered $\ell_\mathrm{max}$.  Next, we consider the change of the method for computing the covariance.  In the next
    row, {\tt Planck} cosmology is assumed for both the BAO template and the
    \cosmolike covariance. Finally, to check the number of broadband
    terms required to get an unbiased estimate for an alternative cosmology
    template, we have shown the results obtained with the {\tt Planck}
    cosmology template for different number of broadband terms. 
    The bold entries indicate the baseline choice, in
    addition to $\ell_{\rm min} =10$ and $\ell_{\rm max}= (410, 470, 510, 550,
    610)$ for each bin (see text for details).}
  \label{tab:Cl_mocktest}
  \begin{tabular}{lccccccccc}
    \hline
    \hline
    case & $\langle \alpha \rangle$ & $\sigma_{\rm std}$ & $\sigma_{68}$ & $\langle \sigma_{\alpha} \rangle$ & ${\rm fraction\,encl.} \langle \alpha \rangle$ & $\langle d_{\rm norm} \rangle$ & $\sigma_{d_{\rm norm}}$ & $\langle \chi^2 \rangle / {\rm d.o.f.}$ & ${\rm mean\,of\,mocks}$ \T\B \\ \hline 
    $i=0$ 		         & 1.006 & 0.020 & 0.019 & 0.019 & 68\% & -0.005 & 1.043 & 118.5/114 (1.04) & 1.006 $\pm$ 0.019 \T \\  
    $i=0,1$		         & 1.002 & 0.023 & 0.021 & 0.022 & 69\% & 0.020  & 1.045 & 113.2/109 (1.04) & 1.002 $\pm$ 0.021    \\
    $i=0,1,2$		         & 1.003 & 0.023 & 0.022 & 0.022 & 68\% & 0.026  & 1.041 & 107.9/104 (1.04) & 1.003 $\pm$ 0.022    \\
    ${\bf i=-1,0,1,2}$       & {\bf 1.004} & {\bf 0.025} & {\bf 0.023} & {\bf 0.023} & {\bf 69\%} & {\bf -0.008} & {\bf 1.050} & {\bf 100.9/99  (1.02)} & {\bf 1.004 $\pm$ 0.023}    \\
      
    $\Delta\ell=10$		 & 1.004 & 0.024 & 0.023 & 0.022 & 67\% & -0.008 & 1.059 & 229.9/226 (1.02) & 1.004 $\pm$ 0.022 \T \\
    ${\bf \Delta\ell=20}$	 & {\bf 1.004} & {\bf 0.025} & {\bf 0.023} & {\bf 0.023} & {\bf 69\%} & {\bf -0.008} & {\bf 1.050} & {\bf 100.9/99  (1.02)} & {\bf 1.004 $\pm$  0.023 }   \\
    $\Delta\ell=30$		 & 1.004 & 0.027 & 0.024 & 0.024 & 68\% & -0.011 & 1.063 & 58.5/57   (1.03) & 1.004 $\pm$ 0.024 \B \\

    $\ell_{\rm max} = 550$  &  1.004  &  0.025  & 0.023   & 0.023  & 67.5\% &
      -0.013 & 1.061 & 112.8/109(1.03) & 1.004 $\pm$ 0.023 \B \\
                                                                                                                                
    ${\rm COLA\,cov}$ 	 & 1.004 & 0.025 & 0.023 & 0.024 & 71\% & -0.010 & 0.974 & 92.6/99   (0.94) & 1.004 $\pm$ 0.023 \T \\
    ${\rm FLASK\,cov}$	 & 1.004 & 0.026 & 0.024 & 0.023 & 67\% & -0.006 & 1.079 & 102.4/99  (1.03) & 1.004 $\pm$ 0.023   \B \\

    ${\rm {\tt Planck}\,cosmology}$ & 0.965 & 0.024 & 0.022 & 0.028 & 78\% & -0.002 & 0.835 & 76.5/99   (0.77) & 0.965 $\pm$ 0.027 \B \\         

    $ {\rm {\tt Planck} \,temp.} \, i=0 $                    & 0.917 & 0.024 & 0.023 & 0.022 & 67\% &  0.101 & 1.076 & 133.2/114 (1.17) & 0.918$\pm$ 0.022 \\
    $ {\rm {\tt Planck} \,temp.} \, i=0, \, 1 $              & 0.947 & 0.025 & 0.023 & 0.023 & 66\% &  0.066 & 1.068 & 121.9/109 (1.12) & 0.948$\pm$ 0.022 \\
    $ {\rm {\tt Planck} \,temp.} \, i=0, \, 1, \, 2 $        & 0.957 & 0.022 & 0.021 & 0.021 & 68\% &  0.044 & 1.046 & 111.3/104 (1.07) & 0.958$\pm$ 0.020 \\
    $ {\rm {\tt Planck} \,temp.} \, i=-1, \, 0, \, 1, \, 2 $ & 0.966 & 0.023 & 0.022 & 0.023 & 70\% & -0.005 & 0.993 & 102.0/99  (1.03) & 0.966$\pm$ 0.022 \B \\

    \hline
  \end{tabular}
\end{table*}

For the {\tt MICE} cosmology template, the fit results are only weakly dependent on the number of braodband terms with the effects on $C_\ell $ being more apparent. We will see below that this is more important when we consider a different cosmology for the template and this will drive the number of broad-band terms used by defaults.  
For $w$, a number of angular bin widths $\Delta \theta $ are considered: $0.10^\circ$, $0.15^\circ$, and  $0.20^\circ $. The results are insensitive to the angular bin width, and we find that the $\chi^2/ {\rm d.o.f.}$ increases mildly with the increase of the bin width.  To reduce the size of the covariance matrix, we adopt  $0.20^\circ $ as the fiducial setup.  Similarly, for $C_\ell $, the results for $\Delta \ell =10$, 20, and 30 are shown. For $\Delta \ell =30$, the difference between $ \sigma_{\rm std }$ and  $ \langle \sigma_{\alpha } \rangle $ is marked, while for smaller $\Delta \ell $, the difference reduces. Again as a compromise for the size of the covariance matrix, we adopt $\Delta \ell = 20$ as the fiducial setup.  Using a minimum angular scale $\theta_{\rm min} = 1^\circ $ for $w$, the results are basically unchanged and this suggests the results are not sensitive to the small scales. 
We conclude that changing the fiducial analysis configuration does not introduce quantitative differences in the results. 

Tables \ref{tab:w_mocktest} and \ref{tab:Cl_mocktest}  also display the results obtained using COLA and FLASK covariance for reference. Overall the COLA covariance gives very consistent results, especially for $w$, e.g.~$ \langle \sigma_\alpha \rangle  $ and  $\sigma_{\rm std}$ (or $\sigma_{\rm 68}$) are closer to each other. This is reassuring since this estimation traces the actual mock statistics. However, the overlapping issue mentioned previously  casts doubt on its validity for the real data. Furthermore, it is only available in the {\tt MICE} cosmology. For the FLASK covariance, we find slightly larger difference between  $ \langle \sigma_\alpha \rangle  $ and  $\sigma_{\rm std}$ (or $\sigma_{\rm 68}$).

The next entries of Tables \ref{tab:w_mocktest} and \ref{tab:Cl_mocktest} show the results obtained assuming a {\tt Planck} cosmology ({\tt Planck} template plus the {\tt Planck} covariance). The theoretically expected value for fitting {\tt MICE} mocks with a {\tt Planck} template (by comparing the sound horizon and the angular diameter distance) is 0.959, if we consider the  0.004 shift found in the mocks due to non-linearities, we are left with an expectation of 0.963. Hence, finding an additional small bias of 0.003 or 0.002 when comparing to the $ \langle \alpha \rangle $ in the {\tt Planck} cosmology entry of Tables \ref{tab:w_mocktest} \& \ref{tab:Cl_mocktest}, respectively. Nevertheless, this bias is negligible when compared to the error bars.
The fact that the magnitude of the covariance elements are generally larger in {\tt Planck} cosmology can explain that $ \sigma_\alpha $ and fraction enclosing  $\langle \alpha \rangle $ are higher while $\chi^2/\mathrm{d.o.f.}$ and $\sigma_{d_{\rm norm} }$ are lower. 
The last set of entries in Tables \ref{tab:w_mocktest} and \ref{tab:Cl_mocktest} consider fitting the COLA mocks (based on MICE cosmology) with the Planck template (but still the MICE covariance) for different number of broad-band parameters. In this case, it is very clear that the number of broad-band parameters is very important to recover unbiased results. For APS, we find that we need at least 4 ($i=-1, 0, 1, 2$) broad-band parameters so that results converge (not shown here, but results are consistent when using more parameters). For the ACF, we find that 2 or 3 parameters are sufficient to get stable results. For this case, we consider 3 parameters ($i=0, 1, 2$) in order to allow for more flexibility.

\begin{table*}[ht]
  \caption{Results of BAO fits in mock catalogs. The results of ACF and APS are combined using two methods, described in Sec.~\ref{sec:comb}.}
      \label{tab:combined_mocktest}
  \begin{tabular}{lccccccccc}
    \hline
    \hline
    case & $\langle \alpha \rangle$ & $\sigma_{\rm std}$ & $\sigma_{68}$ & $\langle \sigma_{\alpha} \rangle$ & ${\rm fraction\,encl.} \langle \alpha \rangle$ & $\langle d_{\rm norm} \rangle$ & $\sigma_{d_{\rm norm}}$ & ${\rm mean\,of\,mocks}$ \T\B \\ \hline 
    $w(\theta)$     & 1.004 & 0.024 & 0.023 & 0.021 & 62\% & -0.024 & 1.112 & 1.004 $\pm$ 0.021 \\  
    $C_\ell$     & 1.004 & 0.025 & 0.023 & 0.023 & 69\% & -0.008 & 1.050 & 1.004 $\pm$ 0.023 \\
    $w(\theta) + C_\ell$  [method 1: {\bf FID}] \ & 1.004 & 0.024 & 0.022 & 0.022 & 67\% & -0.017 & 1.054 & 1.004 $\pm$ 0.022 \\
    $w(\theta) + C_\ell$  [method 2] & 1.004 & 0.023 & 0.022 & 0.021 & 65\% & -0.017 & 1.102 & 1.004 $\pm$ 0.021 \\
    \hline
  \end{tabular}
\end{table*}

As discussed before we expect our fiducial result to come from the combination of ACF and APS. For combining these two highly correlated statistics we put forward two slightly different methods, presented in Sec.~\ref{sec:comb}.  The combined statistic from both combination methods are shown in \autoref{tab:combined_mocktest}. Using method 1, that considers the geometric mean of the individual likelihoods, we find a mean combined value of $\langle \alpha  \rangle = 1.004$ with a mean combined error $\langle \sigma_\alpha \rangle = 0.022$. This mean error coincides perfectly with the $68\%$ distribution of combined best-fit $\alpha$ values and is smaller than the standard deviation $\sigma_{\rm std}$. The pull statistics using $d_{\rm norm}$ agree to within $5\%$ with that corresponding to a unitary Gaussian. For method 2 we first measure the Pearson correlation coefficient from all the mock measurements, yielding $\rho = 0.893$. Using this value we estimate the weight per mock and the combined $\alpha$ statistics as shown in \autoref{tab:combined_mocktest}. Alternatively, we also tried using the weights derived from the mean error of the mocks, yielding very similar statistics (not explicitly shown here). In method 2, we eliminated by default the mocks that have negative values for either of the weights, this is 35\% of the mocks. Given how often this happens, we also tried including those mocks in the ensemble, only resulting in very minor changes to the statistics in \autoref{tab:combined_mocktest}. 
We do not enter into details these alternative methods and we do not include their values in \autoref{tab:combined_mocktest}, as neither of them is chosen as the fiducial option. We leave for Y6 a more detailed exploration of the methods for the combined constraints.

The results using method 2 are very consistent with those from method 1, with a slightly larger deviation from Gaussian statistics. In all, both methods produce very similar final results but we will consider method 1 as our fiducial. The improvement on the measurement error coming from the combination of ACF and APS is of $\sim 5\%$ on the mocks.

\section{Pre-unblinding tests}
\label{sec:blindtest}

In order to avoid confirmation bias, the analysis was performed blind. While blinded, we do not compute $\alpha$ on the final data vector, and do not plot the ACF or APS. Before un-blinding, we require passing a set of tests designed to identify any issues in the analysis without being influenced by confirmation bias. The pre-unblinding tests are the following, 

\begin{enumerate}
	\item{ {\bf Is the BAO detected?} We consider to have a detection if the $1\sigma$ region of posterior of the $\alpha$ parameter (using all 5 tomographic bins) lies inside the prior range $[0.8, 1.2]$, i.e, the $\alpha \pm 1 \sigma_\alpha$ is within our (flat) prior limit. We find this test to pass for the DES Y3 BAO sample data. This interval is 10 times larger than the expected error in $\alpha$, and we can know if the measurement satisfies this constraint without violating the blinding protocol.} 

	\item{ {\bf Is the measurement robust?} We test the impact in $\alpha$ from variations in the analysis. In order to remain blinded we only refer to variations in  $\alpha$ with respect to a fiducial analysis, defining a variable $\Delta  = \alpha_{\rm variation} - \alpha_{\rm fid}$ in each case. We repeat this process on mocks and consider the data to have passed an individual test if we find $\Delta \alpha^{\rm data}$ to be within the $90\%$ confidence interval of $\Delta \alpha^{\rm mock}$. If one or more tests do not pass, we consider the ensemble statistic of all tests to determine the likelihood of such a failure (similarly if two or more tests do not pass, etc). If the ensemble  probability of such a failure is $10\%$ or more we consider the failure statistically justified and move on. If it is less than $10\%$ but larger than $1\%$ we delay the unblinding stage to re-evaluate all the analysis process.  If we find nothing in the process that needs to be changed we move on. If the ensemble probability of such failure is below $1\%$, and remains so after the revision, the unblinding is not allowed.}
The individual tests are the following, 

	\begin{itemize}
		\item{ {\bf Impact of removing one tomographic bin.} We test the change in best-fit $\alpha$ when removing one tomographic bin at a time and compare to the equivalent distribution in the COLA mocks. The quantity being measured on both the mocks and data is $\Delta = \alpha_{4-\rm bins} - \alpha_{5-\rm bins}$.  For the mocks this test is done using a {\tt MICE} cosmology template (corresponding to the true cosmology of the mocks). For the data we perform the test with {\tt Planck} or {\tt MICE} cosmologies. The results are shown in the top five rows of Tables \ref{tab:unblind_aps} and \ref{tab:unblind_acf}. We also test the impact of removing tomographic bins on the estimated uncertainty $\sigma_{\alpha}$, which is displayed in the bottom part of each table. The result on the data agrees well with the distributions on the mocks. While this is not a pre-unblinding requirement in itself we regard it as informative.}

		\item{ {\bf Impact of template cosmology.} We test that our results vary with the assumed cosmology template as expected from the change in cosmology itself. Hence we perform the BAO fits assuming either a {\tt MICE} or {\tt Planck} template (using the same covariance). From these cosmologies we theoretically expect to find a difference in $\alpha$ of 0.041. However, in \autoref{tab:w_mocktest} \& \autoref{tab:Cl_mocktest} we found that, for the mocks, the mean shift of $\alpha$ is 0.038 \& 0.039 for the ACF \& APS statistics, respectively.
		Hence we consider the variable $\Delta = (\alpha_{\rm Planck} - \alpha_{\rm MICE} + 0.039)$, which should be centered around zero for the APS and around 0.001 for the ACF. The results are shown in the sixth row of Tables \ref{tab:unblind_aps} and \ref{tab:unblind_acf} (under column {\tt MICE} for the data).}
		\item{ {\bf Impact of covariance cosmology.} We measure the distribution of best fit $\alpha$ when using \cosmolike covariance with {\tt MICE} cosmology and bias values (our default for these set of tests) or {\tt Planck} cosmology (with its corresponding refitted galaxy bias). The variable here is $\Delta = \alpha_{\rm Planck \ \cosmolike} - \alpha_{\rm MICE \ \cosmolike}$. Results are displayed in the seventh row to Tables \ref{tab:unblind_aps} and \ref{tab:unblind_acf}.}
		\item{ {\bf Impact of n(z) estimation.} We test that our results are robust with respect to the estimation of the underlying redshift distributions. 
		We compare the best fit $\alpha$ when fitting with BAO templates obtained with a redshift distribution from VIPERS direct calibration or using the stacking of \DNF \ZMC, see Sec.~\ref{sec:photo-z} and \cite{y3-baosample}. The variable is $\Delta = (\alpha_{\DNF-\ZMC} - \alpha_{\rm VIPERS})$. We compare this to the same quantity on the COLA mocks. We perform this test for two different cosmologies, and the results are shown in the eighth row of the table.}	
	\end{itemize}
	These tests are summarised in Tables \ref{tab:unblind_aps} and \ref{tab:unblind_acf}. The test removing the 5th redshift bin fails at 90\% Confidence Level (CL) for the power spectrum and at 97\% CL for the correlation function. We also find that the test of the impact of template cosmology is failed at the 90\% level for the correlation function.  All other tests pass. For the power spectrum, 43\% of mocks had one or more test fail at 90\% CL, so we consider that the APS passes very clearly the pre-unblinding tests. For the correlation function 17\% of mocks has one or more test fail at 97\% CL and 22\% of mocks fail 2 tests at 90\% CL. Since more than 5\% of mocks fail the same number of robustness tests, given the CL intervals, we consider this pre-unblinding test to pass\footnote{The failure of the template cosmology test was only identified during the refereeing process after unblinding. However, we have found that, given the criteria defined while blinded, we still pass the global tests.}. Additionally, below, we consider the ensemble of robustness tests all together. 

	\item{ {\bf Is it a likely draw?} We measure the covariance of the $\Delta \alpha$ for each of the above tests from the COLA mocks (i.e. an $8\times8$ covariance given the 8 different tests, estimated from 1952 mocks). We compute the $\chi^{2}$ for each mock $\Delta \alpha$ from this covariance. If $\chi^2_{\rm data} > 99\%$ of the $\chi^2_{\rm mock}$ distribution we consider this test to have failed. If $99\% > \chi^2_{\rm data} > 95\%$ of the  $\chi^2_{\rm mock}$, we consider this to warrant further investigation.  On the real data we find the angular power spectrum $\Delta \alpha$ to have $\chi^2=3.58$, much less than the 95\% threshold of 22.4. The angular correlation function $\Delta \alpha$ has $\chi^2=12.5$, much less than the 95\% threshold of 22.5. 
	}
	\item{ {\bf Is it the BAO a good fit?} We compare the goodness of fit of the $w(\theta)$ fit itself, comparing the $\chi^2$ measured on the data to the same value as the mocks. We again consider a p-value of $<1\%$ to be a failed test and $<5\%$ to warrant further investigation. These $\chi^{2}$ values are shown in Table \ref{tab:baodata}, all the corresponding p-values pass the $1\%$ threshold.
	}

\end{enumerate}

For the pre-unblinding tests that require measuring $\alpha$ on the real data we apply an unknown random offset to all $\alpha$ measurements to keep the true measurement blind (besides considering only $\Delta \alpha$ statistics). Since we passed all the pre-unblinding conditions, we proceeded to the final stage.

\begin{table*}[ht!]
\caption{
Table of pre-unblinding tests for the Angular Power Spectrum from Section \ref{sec:blindtest}, showing the impact of removing individual tomographic bins, of changing the assumed cosmology for the BAO template or the covariance, and of considering an independent estimate of the true redshift distributions. We report variations in $\alpha$ with respect to our fiducial analysis, to keep results blind. The first four columns show the range of $\Delta \alpha $ values measured on the COLA mocks that enclose the fraction of mocks shown at the top of each column. By default the mocks are analysed assuming {\tt MICE} cosmology. The fifth column shows the $\Delta \alpha$ value measured on the data, for two different fiducial cosmologies, {\tt MICE} and {\tt Planck}. A test (each row) is said to have failed if the data value falls outside of the bounds measured on the mocks. For the Angular Power Spectrum, only one test ("1234" removing the fifth redshift bin) fails at the 90\% threshold. Considering the whole ensemble of tests (all rows), we find 43\% of mocks have one or more failed test at the 90\% threshold. Hence this failure was not considered an issue for unblinding. The bottom raws show the impact of removing one bin at a time on the error in $\alpha$ (but this does not impose conditions on unblinding).
\\}
\label{tab:unblind_aps}
\begin{tabular}{l|ll|ll|ll|ll||l|l}
\hline
\hline
Threshold       & \multicolumn{2}{l|}{0.9} & \multicolumn{2}{l|}{0.95} & \multicolumn{2}{l|}{0.97} & \multicolumn{2}{l||}{0.99} & \multicolumn{2}{l}{data} \\ \cline{2-11}
(Fraction of mocks)                &          min &          max &          min &          max &          min&          max &          min&          max & {\tt MICE}       & {\tt Planck}       \\ \hline
        \T        & \multicolumn{8}{c||}{$10^2 (\alpha-\alpha_{\rm fiducial})$}                      \B                                                                                 \\ \cline{2-11}            
Bins\,2345            & -1.58        & 1.73         & -2.22        & 2.18         & -2.65       & 2.73         & -4.33       & 4.46         & 0.92       & 1.17         \\
1345            & -1.8         & 1.88         & -2.36        & 2.43         & -2.84       & 2.84         & -4.35       & 3.68         & -0.62      & -0.46        \\
1245            & -1.87        & 2.07         & -2.55        & 2.76         & -3.04       & 3.49         & -4.26       & 5.92         & -0.34      & -0.26        \\
1235            & -2.01        & 1.84         & -2.77        & 2.39         & -3.29       & 3.06         & -4.44       & 4.62         & -0.84      & -1.02        \\
1234            & \bf{-2.28}   & \bf{1.75}    & -3.13        & 2.35         & -3.49       & 2.78         & -5.14       & 3.77         & 1.58       & \bf{1.83}    \\
{\tt Planck} Template  & -0.66        & 0.72         & -0.88        & 0.87         & -1.08       & 1.07         & -1.48       & 1.37         & 0.10       & -            \\
{\tt Planck} Covariance & -0.41        & 0.47         & -0.53        & 0.59         & -0.59       & 0.74         & -0.85       & 0.9          & -0.16      & -            \\
$n(z)_{\DNF- {\rm Z_{\rm MC}}}$   & -0.73        & 0.12         & -0.87        & 0.21         & -0.96       & 0.29         & -1.13       & 0.46         & -0.13      & -0.19        \\
\hline 
          \T      & \multicolumn{8}{c||}{$(\sigma-\sigma_{\rm All \ Bins})/\sigma_{\rm All \ Bins}$}            \B                                                                                           \\ \cline{2-11}            
Bins\,2345            & -0.05               & 0.30                & -0.09               & 0.38                & -0.12              & 0.44                & -0.18              & 0.59                & 0.00       & 0.05         \\
1345            & -0.06               & 0.31                & -0.08               & 0.38                & -0.12              & 0.44                & -0.18              & 0.62                & 0.09       & 0.10         \\
1245            & -0.04               & 0.37                & -0.08               & 0.47                & -0.10              & 0.52                & -0.18              & 0.82                & 0.10       & 0.05         \\
1235            & -0.06               & 0.37                & -0.1                & 0.44                & -0.13              & 0.56                & -0.19              & 0.75                & 0.23       & 0.08         \\
1234            & -0.06               & 0.36                & -0.1                & 0.44                & -0.12              & 0.50                & -0.19              & 0.69                & 0.19       & 0.09         \\
\hline
\end{tabular}
\end{table*}


\begin{table*}[]
\caption{
As table \ref{tab:unblind_aps} but for the Angular Correlation Function. We find one test ("1234" removing the fifth redshift bin) fails at the 97\% threshold. We find 17\% of mocks have one or more failed test at the 97\% threshold so this was not considered an issue for unblinding. \\}
\label{tab:unblind_acf}
\begin{tabular}{l|ll|ll|ll|ll||l|l}
\hline
\hline
Threshold       & \multicolumn{2}{l|}{0.9} & \multicolumn{2}{l|}{0.95} & \multicolumn{2}{l|}{0.97} & \multicolumn{2}{l||}{0.99} & \multicolumn{2}{l}{data} \\ \cline{2-11}
(Fraction of mocks)   &   min &          max   &          min &          max &          min&          max &          min&          max & {\tt MICE}       & {\tt Planck}       \\ \hline
                \T &   \multicolumn{8}{c||}{$10^2 (\alpha-\alpha_{\rm fiducial})$}  \B                                                                                                     \\ \cline{2-11}            
Bins\,2345            &  -1.60       & 1.84      & -2.27       & 2.42       & -2.76       & 2.92       & -4.20       & 3.78       & 1.12       & 0.92        \\
1345            & -1.80       & 1.84      & -2.22       & 2.36       & -2.76       & 2.92       & -3.84       & 4.30       & -1.24      & -1.08       \\
1245            & -1.92       & 1.99      & -2.52       & 2.68       & -3.27       & 3.00       & -4.81       & 4.64       & -0.68      & -0.26       \\
1235            & -1.88       & 1.84      & -2.60       & 2.34       & -3.11       & 2.87       & -4.15       & 4.02       & -1.44      & -1.02       \\
1234            & -1.95       & 1.68      & -2.56       & 2.26       & \bf{-3.12}  & \bf{2.68}  & -4.15       & 3.68       & \bf{3.44}  & \bf{2.84}   \\
{\tt Planck} Template  & \bf{-0.38}       & \bf{0.54}      & -0.50       & 0.66       & -0.58       & 0.74       & -0.70       & 0.90       & \bf{0.61}       & -           \\
{\tt Planck} Covariance \ \      & -0.40       & 0.36      & -0.52       & 0.48       & -0.60       & 0.55       & -0.77       & 0.76       & 0.08       & -           \\
$n(z)_{\DNF- {\rm Z_{\rm MC}}}$   & -0.80       & 0.12      & -0.92       & 0.20       & -1.00       & 0.31       & -1.21       & 5.21       & 0.00       & -0.12       \\
\hline
            \T    & \multicolumn{8}{c||}{$(\sigma-\sigma_{\rm All \ Bins})/\sigma_{\rm All \ Bins}$}  \B                                                                                                      \\ \cline{2-11}            
Bins\,2345            & -0.04               & 0.30                & -0.06               & 0.36                & -0.10              & 0.42                & -0.14              & 0.62                & 0.05       & 0.05         \\
1345            & -0.04               & 0.33                & -0.06               & 0.40                & -0.09              & 0.47                & -0.15              & 0.58                & 0.04       & 0.05         \\
1245            & -0.04               & 0.37                & -0.07               & 0.45                & -0.09              & 0.54                & -0.16              & 0.74                & 0.12       & 0.11         \\
1235            & -0.05               & 0.35                & -0.08               & 0.43                & -0.11              & 0.53                & -0.11              & 0.53                & 0.09       & 0.12         \\
1234            & -0.05               & 0.35                & -0.07               & 0.40                & -0.09              & 0.45                & -0.14              & 0.70                & 0.07       & 0.07         \\
\hline
\end{tabular}
\end{table*}

\section{Results}
\label{sec:results}

The different methods previously discussed produce BAO detections in both real and harmonic space. We study the compatibility
of these detections and their combination, that yield a precise determination of the angular
diameter distance to $z = 0.835$.

\begin{figure}
  \includegraphics[width=\linewidth]{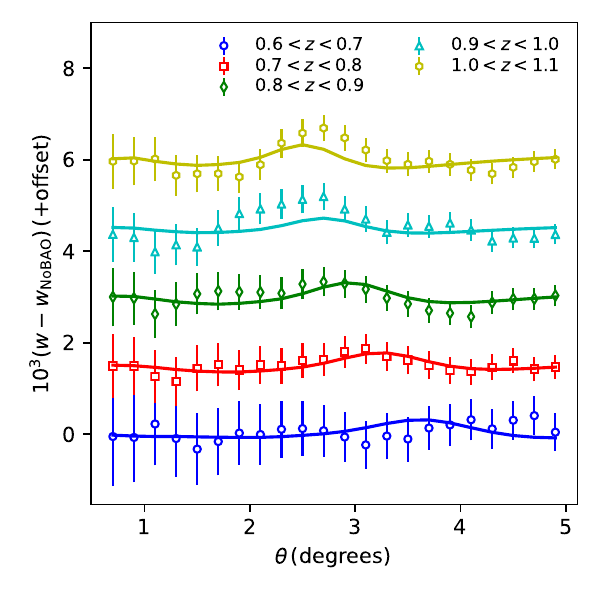}
  \caption{The isolated BAO feature, measured in configuration space using the
    angular correlation function, $w(\theta)$. The measurements have been
    re-scaled by a factor of 10$^{3}$ and vertical offsets of +1.5 have been
    sequentially added with each tomographic bin. The BAO feature moves to
    lower angular scales as the redshift increases, reflecting constant
    co-moving size. The error bars are based on the fiducial Planck 
    covariance, neighboring data points are strongly correlated. The solid lines represent the best fit.}
  \label{fig:ACF-data}
\end{figure}

\begin{figure}
  \includegraphics[width=\linewidth]{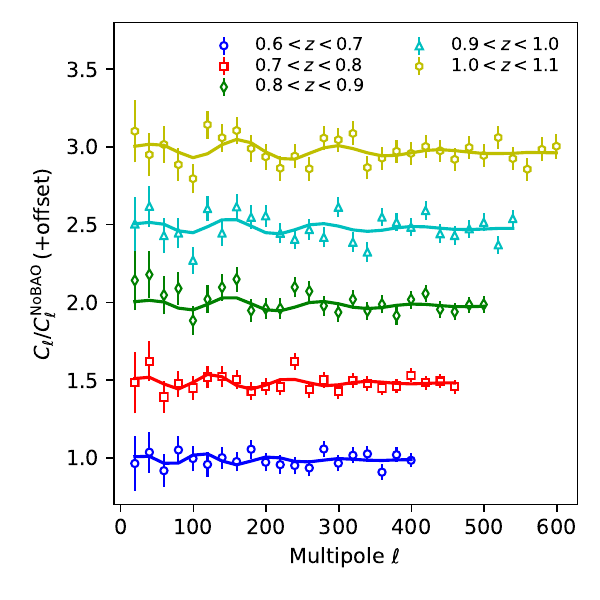}
  \caption{The isolated measured BAO feature, same as
    Figure~\ref{fig:ACF-data}, but in harmonic space, using the angular power
    spectrum, $C_{\ell}$. Vertical offsets of +0.5 have been sequentially added
    with each tomographic bin. The BAO feature stretches to larger values of
    $\ell$ with increasing redshift, as it has the same co-moving scale. The
    error bars are based on the fiducial Planck covariance. The solid lines represent the best fit.}
  \label{fig:APS-data} 
\end{figure}

\begin{figure}
    \centering
    \includegraphics[width=\linewidth]{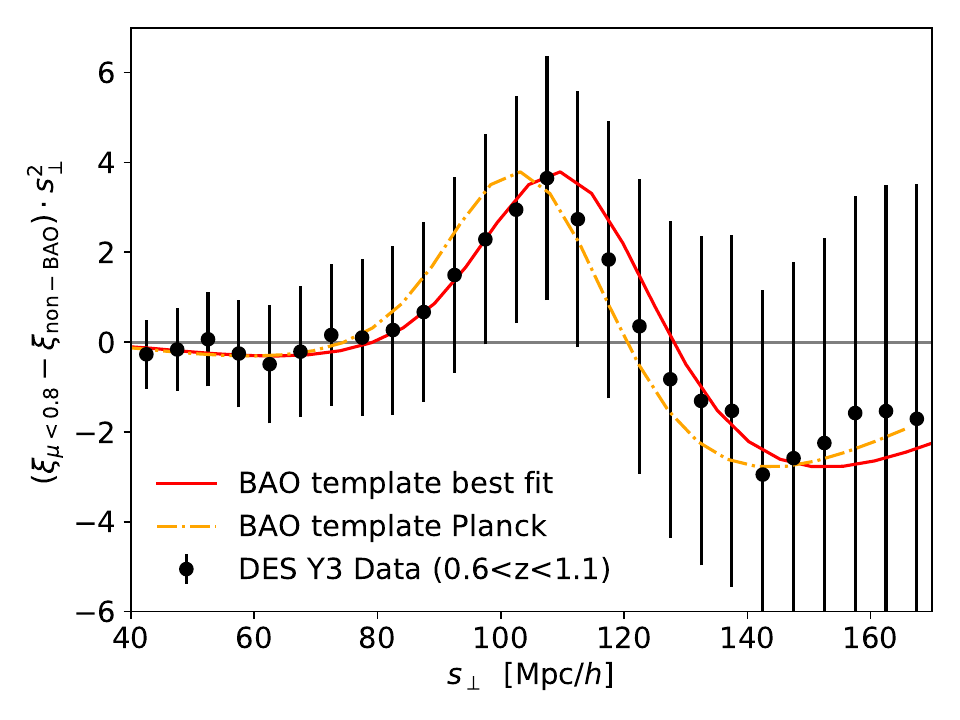}
    \caption{Projected 2-point correlation function as a function of comoving separation perpendicular 
    to the line-of-sight, see Eq.~(\ref{eq:3D}). The circles show the measurement on the DES Y3 
    data with error bars computed from the COLA mocks. The orange dash-dotted line has been shifted by 
    the parameter $\alpha$ corresponding to {\tt Planck} cosmology and red solid line by the $\alpha$ 
    corresponding to the best fit to the $w(\theta)$+$C_\ell$ data. All curves have been subtracted by a 
    non-BAO template. }
    \label{fig:3D}
\end{figure}

\subsection{The BAO signal}

The measured BAO signal is shown in Figure~\ref{fig:ACF-data} for configuration space
and in~\ref{fig:APS-data} for harmonic space. In order to isolate this
signal, we have subtracted (divided) the $w$ ($C_{\ell}$) by the no-BAO smooth
prediction from the no-wiggle power spectrum, Eq.~(\ref{eq:pknw}). To provide better
visualization of each tomographic bin, we further introduced vertical offsets,
displaying the different tomographic bins from the bottom to the top in
increasing redshift order. The BAO feature moves towards lower angular 
scales (lower $\theta$ for $w$ and higher $\ell$ for
$C_{\ell}$) as the redshift increases, reflecting its fixed co-moving scale. 

The projected clustering results can be found in Figure \ref{fig:3D}, where we show 
the measurements on the data following the procedure explained in Sec.~\ref{subsec:projectedclustering}, compared 
to two theoretical curves. First, the {\tt MICE} theory template was created using 
{\tt MICE} cosmology, the biases $b(z_i)$ used for the COLA galaxy mocks and the theory from 
Section~\ref{sec:template} projected to $s_\perp$-space using a framework similar to 
what is described in \cite{Ross:2017emc}. Additionally, a non-BAO template is created 
in a similar fashion for the $w(\theta)$ and $C_\ell$ statistics, which is subtracted from 
all the other curves. We then shift the templates using the $\alpha$ parameter corresponding 
to the {\tt Planck} cosmology and to the best fit BAO resulting from the combination of ACF 
and APS. The templates are rescaled by a factor of  $B^2=0.9$ in order to match the 
amplitude of the DES Y3 clustering at large scales. The errors are computed from the standard 
deviation of 200 COLA mocks, 
and both data and mock pair-counts assume {\tt MICE} cosmology. We only use the projected clustering 
for display purposes since it is the single correlation with highest signal-to-noise, but not 
to derive distance measurements.  

\begin{figure}
  \includegraphics[trim = 0.2cm 0.5cm 0.2cm 0, width= 0.9 \linewidth]{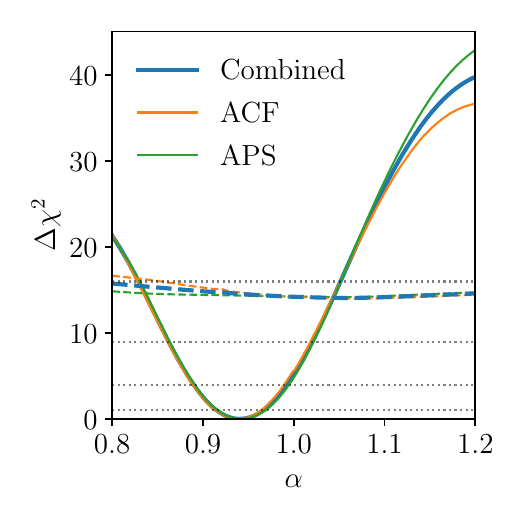}
  \caption{The BAO likelihood for the DES Y3 BAO sample data, for the configuration space (ACF),  the
    harmonic space (APS) and the combined result using the geometric mean of the individual likelihoods. The dashed lines show the recovered likelihood for a
    model without BAO, indicating a detection of the BAO feature in DES Y3 at greater than
    than 3$\sigma$ significance level. The dotted black lines denote 1, 2, 3 and
    4$\sigma$ levels, based on $x\sigma=\sqrt{\Delta \chi^2}$.
    }
  \label{fig:Chi2-data}
\end{figure}

\subsection{BAO Distance Measurements}

Figure~\ref{fig:Chi2-data} shows the $\Delta\chi^2$ for $\alpha$ from the
configuration and harmonic space analyses together with their combination. Both
$\Delta\chi^2$ are remarkably similar, particularly around its minima, diverging 
mostly for large values of $\alpha$. This consistency between the analyses enables a robust combination. In addition, the result for a model
without BAO is also shown, as dashed lines. This model does not describe the
data. From these results, a significance of more than $3 \sigma$ in the 
determination of the BAO feature is inferred.

We find that the values of $\alpha$ measured from these likelihoods in each space, i.e. using the 5 redshift bins together, are all compatible. 
Therefore, we can combine them, obtaining  $\alpha=0.937 \pm 0.025$ with a $\chi^2 = 95.24$ for 89 degrees of freedom and
$\alpha=0.942 \pm 0.026$ with a $\chi^2 = 92.26$ for 99 degrees of freedom, for
the ACF and APS, respectively. These values are in perfect agreement, with 
best-fit values, estimated errors and $\chi^2$ well within the expectations. Thus, we
can also combine the ACF and APS likelihoods, which yields a measurement of $\alpha =
0.941 \pm 0.026$. We consider this our official final measurement. The combined likelihood given by the geometric mean of the individual 
likelihoods is shown by a blue solid line in Fig.~\ref{fig:Chi2-data}.

\begin{figure}
  \includegraphics[width=\linewidth]{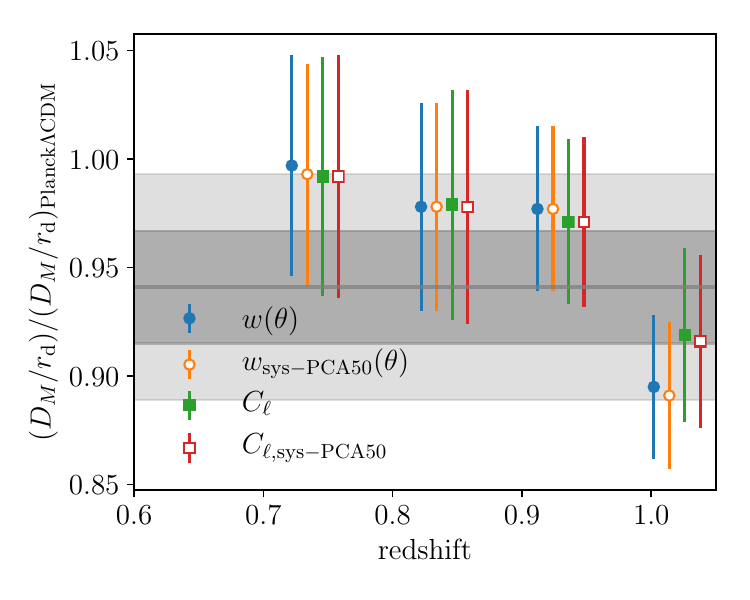}
  \caption{The BAO fit for the individual tomographic bins.
    The shaded region corresponds to our Y3 choice measurement from the
    combination of $w(\theta)$ and $C_{\ell}$.
    We show the individual results for the fiducial analysis and the case of
    using the PCA50 weights.
    The largest differences when changing the weights are for the second and
    fifth bin, being of $\sim 0.1\sigma$ showing the robustness of our
    measurements to systematic weights.
  }
  \label{fig:baofit_onezbin}
\end{figure}

The quality of the Y3 data are high enough to allow us to detect the BAO signals from individual redshift bins.  These are shown in Fig.~\ref{fig:baofit_onezbin}. For $w$, the BAO fit results using individual redshift bins are: No detection, $0.997 \pm 0.051$, $0.978 \pm 0.048$,  $0.977 \pm 0.038$, and  $ 0.895 \pm 0.033 $, with $\chi^2/{\rm dof}$ given by $13.7/17$, $16.6/17$, $23.1/17$ and $10.1/17$. For $C_\ell$ we obtain: No detection, $0.992 \pm 0.055$, $0.979 \pm 0.053$,  $0.971 \pm 0.038$, and  $ 0.919 \pm 0.040$ with $\chi^2/{\rm dof}$ given by $11.9/17$, $13.1/19$, $33.6/21$ and $21.93/24$.

Except for the first bin, which does not meet our detection criteria, BAO signals are detected in individual redshift bins, with the significance increasing with redshift. The fifth bin is the primary driver for the low value of  $\alpha $ compared to {\tt Planck} cosmology. But results are remarkably compatible in real and harmonic space, and the goodness-of-fit is good.  We note that the decrease of the error bar size with redshift is more pronounced than the results from mocks, in which the mean error fluctuates in between 0.4 and 0.5.

\subsection{Robustness Tests}
\label{sec:robustness_tests}

All the robustness tests that have been described for mocks have been also applied to
the data, in order to check the stability and soundness of the results.

First, we have used a second method to combine ACF and APS results, presented in Sec.~\ref{sec:comb}. We assume the
Gaussianity for both likelihoods and then combine the results as correlated 
Gaussian measurements. The correlation between the two $\alpha$ values is obtained
from the mocks. The obtained result (Method 2) is perfectly compatible with the 
fiducial result (Method 1, FID), as can be seen in Table ~\ref{tab:baodata}. 

\begin{table}
\centering
\caption{BAO fits to the Y3 data.  Our combined result is shown in the top row, in terms of the ratio of the recovered comoving angular diameter distance at $z_{\rm eff}=0.835$ to the
  sound horizon scale, for the {\tt Planck} cosmology. The following rows show various robustness tests conducted, discussed in detail in Sec.~\ref{sec:robustness_tests}.}
\begin{tabular}{lcc}
  \hline
  \hline
  {\bf Y3 Measurement} & $D_M/r_{\rm d}$  \T  \\   $z_{\rm eff}=0.835$ & $18.92\pm 0.51$ \B \\
  \hline
  \T case \B & $\alpha$  &  $\chi^2$/dof \\
  \hline
  \T $w(\theta)+C_\ell$ [method 1: {\bf FID}] & $ 0.941 \pm 0.026$   & - \\
  $w(\theta)+C_\ell$ [method 2]        & $ 0.939 \pm 0.025$   & - \B \\
  \hline
  \T Robustness tests: \B \\
  $w(\theta)$                 & $0.937 \pm 0.025$ & $95.2/89$ \\
  $w(\theta)$ no ${\rm sys}$  & $0.935 \pm 0.026$ & $94.6/89$ \\
  $w(\theta)$ ${\rm sys-PCA50}$ & $0.937 \pm 0.025$ & $94.9/89$ \\
  $w(\theta)$ $n(z)$ \DNF\,{\rm PDFs}& $0.935 \pm 0.025$ & $95.6/89$ \\
  $w(\theta)$ $\theta_{\rm min} = 1^{\circ}$  & $0.939 \pm 0.025$ & $81.7/79$ \\
  $w(\theta)$ $\theta_{\rm max} = 4^{\circ}$  & $0.937 \pm 0.025$ & $ 54.7 / 64$ \\
  $w(\theta)$ $\Delta \theta =0.1^{\circ}$   & $0.942 \pm 0.026$   & $ 220.2 / 204$ \\
  $w(\theta)$ 2345 & $0.948 \pm 0.026$ & $67.8/71$ \\ 
  $w(\theta)$ 1345 & $0.929 \pm 0.026$ & $80.7/71$ \\ 
  $w(\theta)$ 1245 & $0.935 \pm 0.028$ & $78.4/71$ \\ 
  $w(\theta)$ 1235 & $0.925 \pm 0.028$ & $70.0/71$ \\ 
  $w(\theta)$ 1234 & $0.967 \pm 0.026$ & $82.3/71$ \\ 
   \T $C_\ell$ & $0.942 \pm 0.026$ & $92.3/99$ \\
  $C_\ell$ no ${\rm sys}$ & $0.940 \pm 0.028$ & $89.7/99$ \\
  $C_\ell$ ${\rm sys-PCA50}$ & $0.941 \pm 0.026$ & $89.7/99$ \\
  $C_\ell$ $n(z) $ \DNF\,{\rm PDFs}& $0.940 \pm 0.025$  & $92.1/99$\\
  $C_\ell$ $\ell_{\rm max}=550$ & $0.940 \pm 0.026$ & $104.6/109$\\
  $C_\ell$ $\Delta\ell=10$ & $0.939 \pm 0.027$ & $238.8/226$ \\
  $C_\ell$ $\Delta\ell=30$ & $0.936 \pm 0.028$ & $40.0/57$ \\
  $C_\ell$ 2345 & $0.954 \pm 0.028$ & $82.2/84$ \\ 
  $C_\ell$ 1345 & $0.938 \pm 0.029$ & $82.0/81$ \\ 
  $C_\ell$ 1245 & $0.940 \pm 0.028$ & $79.5/79$ \\ 
  $C_\ell$ 1235 & $0.932 \pm 0.029$ & $56.3/77$ \\ 
  $C_\ell$ 1234 & $0.961 \pm 0.029$ & $69.1/74$ \B \\ 
  \hline
  \hline
  \label{tab:baodata}
\end{tabular}
\end{table}

Then, we apply the full battery of tests as described below. \vspace{0.2cm}

{\bf $\bullet$ Impact of standard weights for systematics.} The stability of the measurement with respect 
to the observational systematics decontamination has been tested by comparing the
distribution of best fit $\alpha$ values when fitted to mocks uncontaminated by systematics,
contaminated, and decontaminated using our fiducial pipeline. Results of this test are presented in \cite{y3-baosample} 
and indicate the weights have little impact on the
recovered BAO. We compute the same quantity in the data.
    
The BAO fit was found robust to observational systematic effects on the
clustering signal. Even if we do not consider the systematic weights, the BAO
measurement only shifts by $\Delta = 0.002$, i.e., $0.08\sigma$, towards lower
values, both for $w(\theta)$ and $C_{\ell}$. 
Such shifts are fully consistent with the distribution of shifts obtained on lognormal mocks
which are contaminated with systematics, as presented in detail
in Appendix C of \cite{y3-baosample}.\vspace{0.2cm}
Remarkably, we find that the $\chi^2$ does not get worse when not applying the weights. This means that the broad-band terms are able to capture the change of shape induced by the systematics. We checked that 26\% of the mocks contaminated systematics used in \cite{y3-baosample} present better $\chi^2$ than their decontaminated version and we find that the shift in $\chi^2$ is consistent with the distribution found in those mocks. 

{\bf $\bullet$ Impact of PCA weights for systematics.} A second method for 
determining the decontamination weights has been applied, using the principal 
components of the survey properties maps as templates for the 
observational systematics (see~\cite{y3-galaxyclustering}). A
second set of weights (PCA50), comes out from this analysis. The new weights 
have more impact over the clustering amplitude of the second and
fifth redshift bins, but they essentially leave the BAO fit results
unchanged. For the $w(\theta)$ fit, the best fit BAO remains the same as the
default weight case and the $\chi^2$ of the fit shifts slightly from 95.2
to 94.9. For $C_\ell$ there is a slight shift on the best fit $\alpha$
from 0.942 to 0.941, while the recovered error remains the same and the $\chi^2$
also shifts to a lower value from 92.3 to 89.7. Considering the fit to the
individual tomographic bins, the largest shift in the best fit occurs in
the second and fifth bins; nonetheless, the change in $\alpha$ is still
less than $5\times 10^{-3}$ for $w(\theta)$ and $3\times 10^{-3}$ for
$C_{\ell}$, such shifts correspond to $0.1\sigma$ per redshift bin,
Figure~\ref{fig:baofit_onezbin}. The insensitivity to the weight
demonstrates the robustness of the BAO fit results against the
observational systematics. The results are summarized in Table~\ref{tab:baodata}. \vspace{0.2cm}

{\bf  $\bullet$ Impact of removing one z-bin on error bars.} We test the impact of 
removing any tomographic bin on the recovered error bars. The effect is
quantified with the relative deviation $\Delta_{\sigma} = (\sigma-\sigma_{\rm All})/\sigma_{\rm All}$. 
For the $w(\theta)$ we find $\Delta_\sigma=0.05, 0.05, 0.11, 0.12, 0.07$ and for $C_{\ell}$,
$\Delta_\sigma=0.05, 0.10, 0.05, 0.08, 0.09$. We find that $\Delta_{\sigma}$ is always 
positive indicating that the recovered error increases by removing the information from any one bin, and that
none of the bins dominates the total error budget. These values are fully consistent with the distribution of $\Delta_\sigma$ in the mocks, 
shown in the bottom parts of Tables V and VI in Sec.~\ref{sec:blindtest}, even though those correspond to the {\tt MICE} cosmology. \vspace{0.2cm}                 
                                   
{\bf $\bullet$ Impact of  redshift distribution.} In order to test the robustness of  
the BAO measurements to uncertainties in the estimation of the redshift 
distributions, we compare results of a different photo-$z$ method for 
constructing the redshift distributions of the tomographic bins. \red{This test does not change $\ZMEAN$ which is what we use to place galaxies into bins, so the clustering signal itself is unchanged (as well as $z_{\rm eff}$)}. 
The fiducial estimation of the redshift distributions is done by using a calibration 
spectroscopic sample from VIPERS as described in section~\ref{sec:photo-z}. \red{To 
check their robustness we use an alternative determination, from the \DNF 
method \citep{DNF},  which is the stacking of the photo-$z$ PDF per galaxy in each tomographic bin (see \cite{y3-baosample} for further details). Note that VIPERS is not included in the training sample of \DNF so these are two totally independent estimations of the redshift distributions}. This test effectively changes the fiducial templates used for the analysis. We quantify the absolute shift on the BAO measurements via
$\Delta=\alpha_{\DNF\,{\rm PDFs}}-\alpha_{\rm VIPERS}$, obtaining a shift of
$\Delta=0.002$, i.e., $0.08\sigma$ both for $w(\theta)$ and for the 
$C_{\ell}$. A further estimate of the $n(z)$ is to stack the DNF $\ZMC$ values, which yields similar conclusions. These results are also perfectly consistent with the mocks determination studied in Sec. \ref{sec:blindtest} and shown in Tables \ref{tab:unblind_aps} and \ref{tab:unblind_acf}.

The above test probes the ensemble sensitivity of our BAO measurement to uncertainties in our estimation of the true redshift distributions.

In turn, Fig.~\ref{fig:baofit_onezbin} shows that while the second, third, and fourth 
bins give results more or less consistent with each other, the result for the 
last bin is lower, driving the low value of $\alpha $ we find compared to the 
fiducial one. Since the high redshift bin is more prone to photo-$z$ error, it is 
instructive to estimate the error in the photo-$z$ that would result in such a 
shift. We can easily estimate this in configuration space \citep{Chan:2018gtc}.  For 
example, suppose that the deviation of the $\alpha$ value for the last bin 
from the preceding ones, whose combined best fit is 0.967, is due only to 
photo-$z$ error in the last bin. We assume that the photo-$z$ error manifests as 
a shift in the mean of the redshift distribution. This shift modifies the 
comoving angular diameter distance $D_M$ in 
Eq.~(\ref{eq:alpha_BAOtransverse}), and we have
\be
 \frac{ \Delta D_{M}}{  D_{M} }=  - \frac{\Delta \alpha}{ \alpha } . 
\ee
\red{Using this relation and Eq. (29) in \cite{Chan:2018gtc} we can link a shift $\Delta z$ in the mean of a redshift distribution to a change in $D_{M}$ as $\Delta D_{M} / D_{M} \approx - c \, \Delta z / (D_{\rm M}(z) H(z)) $. In our case, the BAO fit to the fifth bin only yields $\alpha=0.895$.  In order to bring this to match the results from the first four bins combined we need $ \Delta D_{M} / D_{M} = - 0.08$,  which amounts to a mean redshift error of $\Delta z = -0.11$.}
\red{We have checked explicitly that this estimate is in very good agreement 
with direct numerical fits obtained  by shifting the actual photo-$z$ 
distribution}. In Fig.~3 of \cite{y3-baosample} we show the degree on uncertainty on the mean of the redshift distributions estimated from \DNF and VIPERS. A value of $\Delta z \sim - 0.11 $ is at least five times larger than the typical photo-$z$ uncertainty of the sample at the last bin. 

\red{Given the difference in the recovered BAO results from all bins is less that 0.1$\sigma$, and that photo-$z$ uncertainties do not seem to be driving the results on the last tomographic bin,  we consider the impact of systematic uncertainties in the $n(z)$ estimation to be negligible for our Y3 BAO measurement.}

\section{Cosmological Implications}
\label{sec:cosmo}

\begin{figure}
  \centerline{\includegraphics[width=\linewidth]{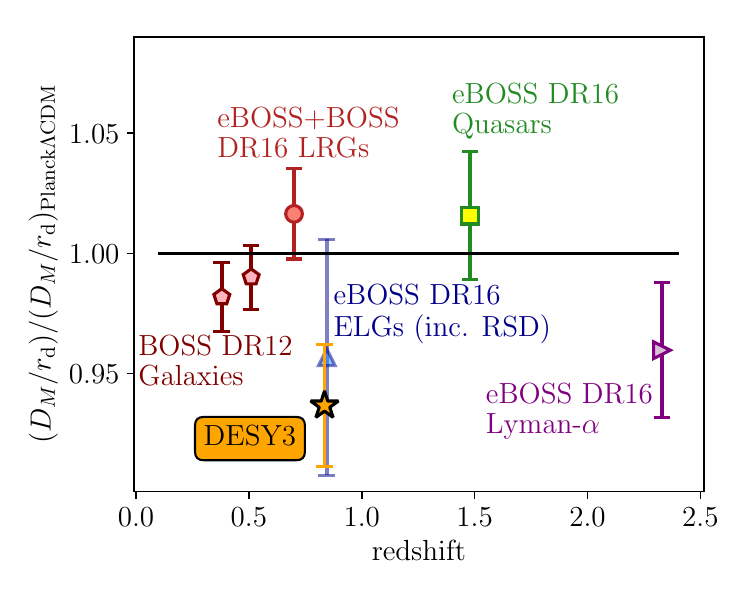}}
  \caption{Ratio between the angular diameter distances measured using the BAO
  feature at different redshifts for several galaxy surveys and the prediction 
  from the cosmological parameters determined by Planck. The DES Y3 measurement is shown by a golden star.}
  \label{fig:baomeasurements}
\end{figure}

\begin{figure}
  \centerline{\includegraphics[width=0.93\linewidth]{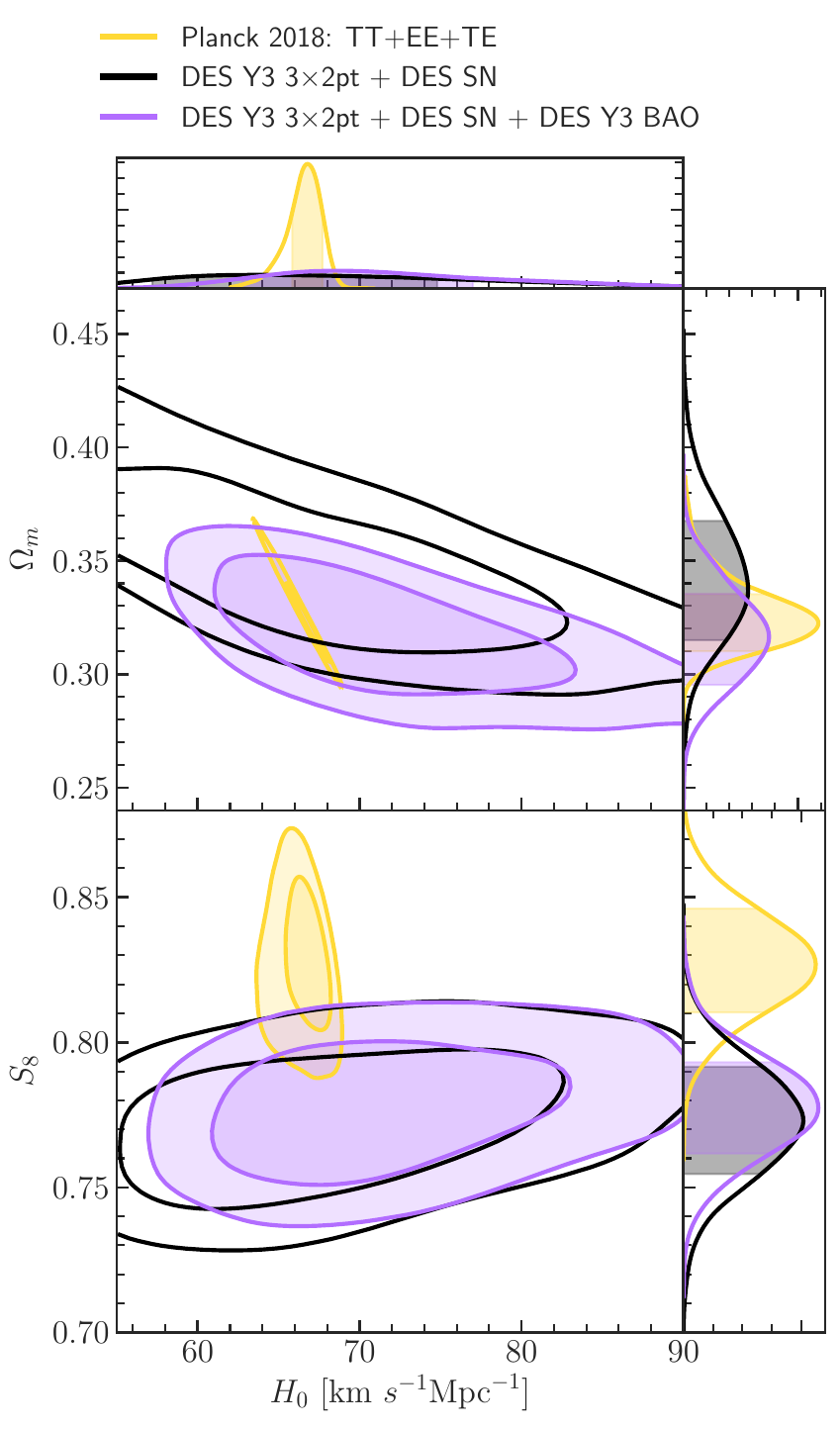}}
  \caption{Marginalised constraints in $\Omega_m$, $S_8$ and $H_0$ from different DES datasets: the combination of clustering and lensing (3$\times$2pt), Supernovae (SNe) and BAO. It highlights the degeneracy breaking introduced by BAO enabling to improve constraints on $\Omega_m$ and particularly $H_0$. The Planck contours are shown in yellow, with $H_0$ and $\Omega_m$ posteriors being fully consistent, and $S_8$ being $\lesssim 1.5\sigma$ low \cite{y3-3x2ptkp}.}
  \label{fig:3x2-SN-BAO}
\end{figure}
 
We now discuss the cosmological implications of our results. Using the Planck cosmology 
and its measurement of the sound horizon at the drag epoch we can convert our best fit $\alpha$ 
in Table \ref{tab:baodata} into $D_M / r_{\rm d}$ measurements at our effective redshift, see Eq.~(\ref{eq:alpha_BAOtransverse}). 
We obtain $D_M / r_{\rm d} = 18.92 \pm 0.51$  at $z_{\rm eff} = 0.835$ from the combination of $w(\theta)$ and $C_\ell$ statistics.  
This represents a fractional error of $2.7\%$, the smallest from any BAO measurement from purely photometric data so far. 

Figure~\ref{fig:baomeasurements} displays the comoving angular diameter distance measurement from the DES Y3 BAO sample normalised by the sound horizon scale (golden star) compared to other $D_M / r_{\rm d}$ in the literature, and the prediction from Planck \lcdm   \cite{2020A&A...641A...6P} assuming a flat \lcdm cosmological model (with fixed neutrino mass). We include measurements from the combined BOSS LOWz + CMASS galaxy samples (at $0.2 < z < 0.5$ and $0.4 < z < 0.6$) \cite{2017MNRAS.470.2617A}, from the eBOSS Luminous Red galaxies (LRG, $0.6 < z < 1.0$) \cite{2020MNRAS.500..736B,2020MNRAS.498.2492G} and the eBOSS Emission Line galaxies (ELG, $0.6 < z < 1.1$) \cite{2020MNRAS.499.5527T,2021MNRAS.501.5616D}, as well as from the eBOSS Quasars ($0.8 < z < 2.2$) \cite{2020MNRAS.499..210N,2021MNRAS.500.1201H} and the Lyman-$\alpha$ combination of auto-correlation and cross-correlation with quasars ($z > 2.1$) \cite{2020ApJ...901..153D}. These measurements and their cosmological implications are compiled and discussed in \cite{2017MNRAS.470.2617A} and \cite{2021PhRvD.103h3533A}, and represent the most updated BAO distance ladder. 

The DES Y3 BAO distance measurement is lower than the Planck \lcdm prediction by $2.27\sigma$. This value is lower, and represents a more significant difference, than the other BAO measurements in Fig.~\ref{fig:baomeasurements}, but is still statistically consistent with Planck. Roughly 1$\sigma$ of this difference is driven by the last redshift bin, as shown in Table \ref{tab:baodata}.

Lastly, we compare with the recent cosmology result from DES Y3 using the combination of clustering and lensing (a.k.a. "DES 3$\times$2pt."), which is based upon the same \gold catalogue as this paper. Converting the 3$\times$2pt \lcdm Monte Carlo Markov Chains (MCMC) from \cite{y3-3x2ptkp} into a posterior distribution on $D_M / r_{\rm d}$ we obtain, $D_M(z_{\rm eff} =0.835)/r_{\rm d} = 20.139 \pm 0.69$ which is bigger than the DES BAO result by about $1.4\sigma$ (accounting for the joint $68\%$ confidence level); while consistent with the Planck result of $D_M(z_{\rm eff} =0.835)/r_{\rm d} = 20.1$. 

The relevance of the DES Y3 BAO measurement for the DES experiment itself is given in Fig.~\ref{fig:3x2-SN-BAO}, that shows marginalised contours on cosmological parameters assuming flat \lcdm from the combinations of DES 3$\times$2pt \cite{y3-3x2ptkp}, Supernovae (SNe) \cite{2019ApJ...872L..30A} and BAO, in addition to Planck \cite{2020A&A...641A...6P}. Details of the DES likelihoods, scale cuts and priors are detailed in \cite{y3-3x2ptkp,2019ApJ...872L..30A} and references therein. DES 3$\times$2pt can constrain well the parameters $\Omega_m$ (matter density) and $\sigma_8$ (amplitude of density fluctuations), and in particular the combination $S_8 \equiv \sigma_8 (\Omega_m / 0.3)^2$, but not the Hubble constant $H_0 \equiv 100 h \, {\rm km~s^{-1}} {\rm Mpc}^{-1} $. The constraining power in $\Omega_m$ is further increased by the inclusion of DES SNe \cite{2019ApJ...872L..30A}, as discussed in \cite{y3-3x2ptkp}. 
Including a BAO measurement helps constraining the absolute calibration of SNe that propagates to a better measurement of $H_0$, in consequence shrinking the ($\Omega_m , h)$ and $(S_8, h)$ contours shown in Fig.~\ref{fig:3x2-SN-BAO}.
We obtain for DES 3$\times$2pt + DES SNe, 
\begin{eqnarray}
h & = &  0.691^{+0.138}_{-0.043}, \\ 
\Omega_m & = & 0.344^{+0.029}_{-0.025}, \\
S_8 & = & 0.773^{+0.018}_{-0.019},
\end{eqnarray}
and for DES 3$\times$2pt + DES SNe + DES BAO,
\begin{eqnarray}
h  & = & 0.72^{+0.090}_{-0.053}, \\
\Omega_m & = & 0.317^{+0.021}_{-0.020},\\
S_8 & = & 0.778^{+0.016}_{-0.017}.
\end{eqnarray}
While the central values are consistent, the posterior in $H_0$ is made more symmetrical by penalizing low $H_0$ values (the prior on $h$ is flat in $0.55-0.91$), with a gain in constraining power of $\sim 20\%$. In turn, the error in $\Omega_m$ is reduced by $\sim26\%$ \footnote{These results neglect any potential covariance between BAO and 3x2pt since the two sets of measurements exploit different redshift ranges and comoving scales, as discussed in \cite{y3-3x2ptkp}.}. The constraining power in $S_8$ improves by $\sim12\%$ with the inclusion of DES BAO.

\section{Conclusions}
\label{sec:conclusions}

We have measured the ratio of angular diameter distance at an effective 
redshift $z_{\rm eff} = 0.835$  to the physical acoustic scale to be $D_M(z_{\rm eff}=0.835)/r_{\rm d} = 18.92\pm 0.51$. These results 
are consistent with the flat \lcdm cosmological model to a precision of $2.7 \%$. We
used clustering measurements from an optimised galaxy sample built upon the first three years of DES data (DES Y3) and described in detail in \cite{y3-baosample}. The sample amounts to 7 million galaxies that cover 
4100 deg$^2$ on the sky in the redshift range $0.6 < z_{\rm photo} < 1.1$. This result is consistent with our measurement using 1500 deg$^2$ from DES Y1 data \cite{2019MNRAS.483.4866A} 
using a similarly defined sample, but improves its precision by $\sim 50\%$ due mainly to the area increase.

The measurement was performed in a fully blinded way to avoid any observer confirmation 
bias. A full battery of quality and consistency tests was defined pre-unblinding and calibrated
with a set of 1952 mock realisations of the DES Y3 BAO sample~\citep{y3-baomocks}. The analysis
choices were fixed using these mocks realiztions as well. Only when the full set of quality and
consistency tests was passed by the data we opened the box to see the results.

Two different determinations of the clustering statistics were used to measure the BAO
scale: angular correlation function $w(\theta)$ (ACF) and  angular power spectrum in spherical 
harmonics, $C_\ell$ (APS). These give very consistent results, as expected from the mock analysis, and presented in Table~\ref{tab:baodata}. The error from APS is only marginally worse by $4\%$ to that from ACF. Moreover, the ACF fit yields a $\chi^2/{\rm dof} = 95.2/89$ corresponding to a $p$-value of 0.3, while the APS yields a $\chi^2/{\rm dof} = 93.2/99$ corresponding to a $p$-value of 0.6. Thus both represent a good fit to the data and we decided to consider the fiducial DES Y3 result to be the
combination of ACF and APS. The BAO feature is detected in the individual and combined likelihoods to a confidence level of more than $3 \sigma$. 

Our result is lower and in some tension with respect to the prediction from Planck assuming a flat \lcdm model, at the level of $2.3\sigma$.
Part of this tension originates from our distance measurement at the highest redshift bin, $1.0 < z_{\rm photo} < 1.1$, both for real and harmonic space.
We have checked that the effects on the measurement due to photometric redshifts uncertainties
and to the correction of survey properties maps with spatial dependence are very well below the
statistical error and have no impact in the final determination. Removing the last bin does not improve the goodness-of-fit of the data or change the fractional errors. Extensive testing with the mocks indicate that the difference in angular diameter distance measurement between the last bin and the ones at lower redshift, \red{while slightly significant, is fully consistent with zero from the statistical point of view given the thresholds that we have defined pre-unblinding}. Summarizing, we have verified that the measurement of the BAO scale in the DES Y3 data is robust
to all the tests we have performed and is consistent with the simulations.

In addition a number of improvements with respect to our previous analysis have been achieved, most notably
\begin{itemize}
 \item{We developed an end-to-end pipeline to produce mock galaxy catalogues automatically matching input number densities, photometric redshift probabilities ${\rm Prob}(z_{\rm phot},z_{\rm spec})$ and galaxy clustering. The pipeline iteratively populates halo light-cone simulations with galaxies following a Halo Occupation Distribution (HOD) prescription, measures the aforementioned quantities to construct a likelihood, and runs a $\chi^2$ minimisation process until it achieves a local minimum. This is fully discussed in \cite{y3-baomocks}.}
 \item{Contrary to previous work, the damping in the BAO template was derived using non-perturbative analytical expressions following \cite{2015JCAP...02..013S,2016JCAP...07..028B}. We tested this yields unbiased results on mocks, equivalent to allowing the damping as a free fitting parameter.  This enabled us to construct templates for arbitrary cosmologies, without the need to prior mock calibration.}
 \item{We have benchmarked with mocks (FLASK, COLA), and finally adopted as default, a halo-model based analytical covariance, \cosmolike \cite{cosmolike,cosmolike2020,cosmolike_curvedsky}. It includes mask effects and higher-order non-gaussian contributions. This enabled us to investigate different assumed cosmologies, besides the one of the mocks.}
\item{We have devised a careful protocol of pre-unblinding checks. From a methodological standpoint, this helps towards converting such protocols in new standards of future BAO analysis.}
\end{itemize}

The BAO angular distance measurement presented in this paper is the most precise measurement in the redshift range $0.6 \lesssim z \lesssim 1.1$, and is competitive with other measurements from spectroscopic  datasets. Together with results from BOSS galaxies, and eBOSS galaxies, quasars and Lyman-$\alpha$ it helps  to construct the most-up-date distance ladder from low to high redshifts ($0.6 \lesssim z \lesssim 2.5$). Our results will be revised with the release of the final DES dataset, using approximately the same area but deeper photometry that will likely enable a higher mean redshift. 

We expect our work to set the stage and motivate future BAO measurements from much larger imaging  datasets, such as those expected from Euclid \cite{2011arXiv1110.3193L}  and LSST \cite{2009arXiv0912.0201L}. These should be able to achieve a percentage level fractional errors, and be complementary to results from DESI \cite{2016arXiv161100036D,2016arXiv161100037D} and Euclid spectroscopy \cite{2011arXiv1110.3193L}.

\section*{Acknowledgments}

Funding for the DES Projects has been provided by the U.S. Department of Energy, the U.S. National Science Foundation, the Ministry of Science and Education of Spain, the Science and Technology Facilities Council of the United Kingdom, the Higher Education Funding Council for England, the National Center for Supercomputing Applications at the University of Illinois at Urbana-Champaign, the Kavli Institute of Cosmological Physics at the University of Chicago, the Center for Cosmology and Astro-Particle Physics at the Ohio State University,the Mitchell Institute for Fundamental Physics and Astronomy at Texas A\&M University, Financiadora de Estudos e Projetos, Funda{\c c}{\~a}o Carlos Chagas Filho de Amparo {\`a} Pesquisa do Estado do Rio de Janeiro, Conselho Nacional de Desenvolvimento Cient{\'i}fico e Tecnol{\'o}gico and the Minist{\'e}rio da Ci{\^e}ncia, Tecnologia e Inova{\c c}{\~a}o, the Deutsche Forschungsgemeinschaft and the Collaborating Institutions in the Dark Energy Survey. 

The Collaborating Institutions are Argonne National Laboratory, the University of California at Santa Cruz, the University of Cambridge, Centro de Investigaciones Energ{\'e}ticas, Medioambientales y Tecnol{\'o}gicas-Madrid, the University of Chicago, University College London, the DES-Brazil Consortium, the University of Edinburgh, the Eidgen{\"o}ssische Technische Hochschule (ETH) Z{\"u}rich, 
Fermi National Accelerator Laboratory, the University of Illinois at Urbana-Champaign, the Institut de Ci{\`e}ncies de l'Espai (IEEC/CSIC), 
the Institut de F{\'i}sica d'Altes Energies, Lawrence Berkeley National Laboratory, the Ludwig-Maximilians Universit{\"a}t M{\"u}nchen and the associated Excellence Cluster Universe, the University of Michigan, NSF's NOIRLab, the University of Nottingham, The Ohio State University, the University of Pennsylvania, the University of Portsmouth, SLAC National Accelerator Laboratory, Stanford University, the University of Sussex, Texas A\&M University, and the OzDES Membership Consortium.

Based in part on observations at Cerro Tololo Inter-American Observatory at NSF's NOIRLab (NOIRLab Prop. ID 2012B-0001; PI: J. Frieman), which is managed by the Association of Universities for Research in Astronomy (AURA) under a cooperative agreement with the National Science Foundation.

The DES data management system is supported by the National Science Foundation under Grant Numbers AST-1138766 and AST-1536171.
The DES participants from Spanish institutions are partially supported by MICINN under grants ESP2017-89838, PGC2018-094773, PGC2018-102021, SEV-2016-0588, SEV-2016-0597, and MDM-2015-0509, some of which include ERDF funds from the European Union. IFAE is partially funded by the CERCA program of the Generalitat de Catalunya. Research leading to these results has received funding from the European Research Council under the European Union's Seventh Framework Program (FP7/2007-2013) including ERC grant agreements 240672, 291329, and 306478. We  acknowledge support from the Brazilian Instituto Nacional de Ci\^encia e Tecnologia (INCT) do e-Universo (CNPq grant 465376/2014-2).

This manuscript has been authored by Fermi Research Alliance, LLC under Contract No. DE-AC02-07CH11359 with the U.S. Department of Energy, Office of Science, Office of High Energy Physics.

\bibliography{refs}

\end{document}